\documentclass[conference]{IEEEtran}
\usepackage[utf8]{inputenc}
\usepackage[pdftex]{graphicx}
\usepackage{soul}
\usepackage{epstopdf}
\usepackage{multirow}
\usepackage{amsmath}
\usepackage{mathtools}
\usepackage{authblk}
\usepackage{cite}
\usepackage{cleveref}
\usepackage{latexsym}
\usepackage[justification=centering]{caption}
\usepackage{caption}
\usepackage{amsthm}
\usepackage{balance}
\usepackage{relsize}
\usepackage{color}
\usepackage{amssymb}
\usepackage{eucal}
\usepackage{url}
\usepackage{comment}
\usepackage{booktabs,subcaption,amsfonts,dcolumn}
\usepackage[nodisplayskipstretch]{setspace}
\usepackage{lipsum}
\usepackage{acronym}
\usepackage{makecell}
\usepackage{mathrsfs}
\usepackage{dsfont}
\usepackage{longtable}
\usepackage{mdframed}
\usepackage{fancybox}

\usepackage
[
        left=1.62cm,
        right=1.62cm,
        top=0.7in,
]{geometry}

\newcolumntype{L}{>{\centering\arraybackslash}m{6cm}}

\newacro{kpi}[KPI]{Key Performance Indicator}
\newacro{tdoc}[Tdoc]{Technical Documents}
\newacro{cr}[CR]{Change Request}

\newacro{rf}[RF]{Radio Frequency}
\newacro{ran}[RAN]{Radio Access Network}
\newacro{sa}[SA]{system architecture}
\newacro{ct}[CT]{core network and terminals}
\newacro{qos}[QoS]{quality of service}
\newacro{wg}[WG]{working group}
\newacro{5ga}[5G-A]{5G advanced}
\newacro{bpe}[BPE]{byte pair encoding}
\newacro{mmb}[MMB]{multi-modal beamforming}
\newacro{jscc}[JSCC]{joint source-channel coding}
\newacro{sft}[SFT]{Supervised Fine-Tuning}
\newacro{dpo}[DPO]{Direct Preference Optimization}
\newacro{llm}[LLM]{Large Language Model}
\newacro{lmm}[LMM]{Large Multi-modal Model}
\newacro{fm}[FM]{Foundation Model}
\newacro{ai}[AI]{Artificial Intelligence}
\newacro{lm}[LM]{Language Modeling}
\newacro{ptlm}[PTLM]{Pre-Trained Language Model}
\newacro{nlp}[NLP]{Natural Language Processing}
\newacro{dl}[DL]{Deep Learning}
\newacro{nn}[NN]{Neural Network}
\newacro{dnn}[DNN]{Deep Neural Network}
\newacro{cnn}[CNN]{Convolutional Neural Network}
\newacro{rnn}[RNN]{Recurrent Neural Network}
\newacro{gnn}[GNN]{Graph Neural Network}
\newacro{ml}[ML]{Machine Learning}
\newacro{cv}[CV]{Computer Vision}
\newacro{ssl}[SSL]{Self-Supervised Learning}
\newacro{tl}[TL]{Transfer Learning}
\newacro{nlm}[NLM]{Neural Language Model}
\newacro{lstm}[LSTM]{Long Short-Term Memory}
\newacro{gpt}[GPT]{Generative Pre-trained Transformer}
\newacro{bert}[BERT]{Bidirectional Encoder Representation from Transformer}
\newacro{nlu}[NLU]{Natural Language Understanding}
\newacro{nlg}[NLG]{Natural Language Generation}
\newacro{t5}[T5]{Text-to-Text Transfer Transformer}
\newacro{icl}[ICL]{In-Context Learning}
\newacro{rlhf}[RLHF]{Reinforcement Learning with Human Feedback}
\newacro{mha}[MHA]{Multi-Head Attention}
\newacro{clm}[CLM]{Causal Language Modeling}
\newacro{mlm}[MLM]{Masked Language Modeling}
\newacro{plm}[PLM]{Permuted Language Modeling}
\newacro{dae}[DAE]{Denoising AutoEncoder}
\newacro{rf}[RF]{Radio Frequency}
\newacro{sota}[SOTA]{state of the art}
\newacro{rag}[RAG]{Retrieval Augmented Generation}
\newacro{moe}[MoE]{Mixture of Expert}
\newacro{peft}[PEFT]{Parameter-Efficient Fine-Tuning}
\newacro{sdo}[SDO]{Standards Developing Organization}
\newacro{cot}[CoT]{Chain-of-Thought}
\newacro{rl}[RL]{Reinforcement Learning}
\newacro{vlm}[VLM]{Visual Language Model}
\newacro{6g}[6G]{Sixth Generation}
\newacro{cv2x}[CV2X]{Cellular Vehicle-to-Everything}
\newacro{esti}[ESTI]{European Telecommunication Standards Institute}
\newacro{oran}[O-RAN]{Open Radio Access Network}
\newacro{qos}[QoS]{Quality of Service}
\newacro{3gpp}[3GPP]{Third Generation Partnership Project}
\newacro{itu}[ITU]{International Telecommunication Union}
\newacro{ran}[RAN]{Radio Access Network}
\newacro{bs}[BS]{Base Station}
\newacro{its}[ITS]{Intelligent Transport System}
\newacro{rrm}[RRM]{Radio Resource Management}
\newacro{lora}[LoRA]{Low Rank Adaptation}
\newacro{mlp}[MLP]{Multi-Layer Perceptron}
\newacro{vit}[ViT]{Vision Transformer}
\newacro{qat}[QAT]{Quantization Aware Training}
\newacro{ptq}[PTQ]{Post-Training Quantization}
\newacro{kv}[KV]{Key-Value}
\newacro{rleif}[RLEIF]{Reinforcement Learning from Evol-Instruct Feedback}
\newacro{v2x}[V2X]{Vehicle to Everything}
\newacro{rag}[RAG]{Retrieval Augmented Generation}
\newacro{fim}[FIM]{Fill-In-the-Middle}
\newacro{mcq}[MCQ]{Multiple-Choice Question}
\newacro{qa}[QA]{Question Answering}
\newacro{ieee}[IEEE]{Institute of Electrical and Electronics Engineers}
\newacro{urllc}[URLLC]{Ultra Reliable and Low Latency Communication}
\newacro{kl}[KL]{Kullback-Leibler}
\newacro{cdf}[CDF]{Cumulative Density Function}
\newacro{rrm}[RRM]{Radio Resource Management}

\title{TelecomGPT: A Framework to Build Telecom-Specfic Large Language Models}

\author[1]{Hang Zou}
\author[1]{Qiyang Zhao}
\author[1]{Yu Tian}
\author[2]{Lina Bariah}
\author[1]{Faouzi Bader}
\author[1]{Thierry Lestable}
\author[1,2]{Merouane Debbah}
        
\affil[1]{Technology Innovation Institute, 9639 Masdar City, Abu Dhabi, UAE }
\affil[2]{Khalifa University, Abu Dhabi 127788, UAE}

\begin{document}

\maketitle

\begin{abstract}
\acfp{llm} have the potential to revolutionize the \acf{6g} communication networks. However, current mainstream \acp{llm} generally lack the specialized knowledge in telecom domain. In this paper, for the first time, we propose a pipeline to adapt any general purpose \acp{llm} to a telecom-specific \acp{llm}. We collect and build telecom-specific pre-train dataset, instruction dataset, preference dataset to perform continual pre-training, instruct tuning and alignment tuning respectively. Besides, due to the lack of widely accepted evaluation benchmarks in telecom domain, we extend existing evaluation benchmarks and proposed three new benchmarks, namely, Telecom Math Modeling, Telecom Open QnA and Telecom Code Tasks. These new benchmarks provide a holistic evaluation of the capabilities of \acp{llm} including math modeling, Open-Ended question answering, code generation, infilling, summarization and analysis in telecom domain.  Our fine-tuned \ac{llm} TelecomGPT outperforms \acf{sota} \acp{llm} including GPT-4, Llama-3 and Mistral in Telecom Math Modeling benchmark significantly and achieve comparable performance in various evaluation benchmarks such as TeleQnA, 3GPP technical documents classification, telecom code summary and generation and infilling. 
\end{abstract}


\section{Introduction}

Recent advances of \acfp{llm} have attracted significant attention across different domains including telecommunications community. \acp{llm}, such as GPT-4 \cite{Achiam2023GPT4TR}, the Llama series \cite{Touvron2023LLaMA, Touvron2023Llama2, llama3}, the Mistral series \cite{Jiang2023Mistral7b, Jiang2024MixtralOE}, and the Falcon series \cite{Almazrouei2023falcon}, have demonstrated remarkable capabilities in both natural language understanding and generation tasks. These \acp{llm} have the potential to revolutionize the 6G communication systems. Unlike conventional \acf{dl} techniques which are already widely applied in telecom domain, \acp{llm} offer enhanced generalization capabilities and the emerging abilities, making them suitable tools for applications such as \acf{rrm} in \acf{oran}.   
Additionally, LLMs can facilitate more sophisticated and adaptive communication protocols, thereby enhancing the efficiency and resilience of telecom networks.

However, integrating mainstream \acp{llm} into current telecom systems presents various challenges. Firstly, mainstream \acp{llm} possess a large model size, hindering their direct deployment in wireless networks. Model compression techniques such as LLM-QAT \cite{liu2023llmqat}, one bit quantization in BitNet \cite{wang2023bitnet}  makes it possible to deploy mainstream \acp{llm} on edge devices, e.g., mobile phones. Second, the long inference time of \acp{llm} is unbearable to meet the requirement of \acf{urllc} in beyond 5G networks. Taking the example of \acf{v2x} communication networks, it would be impossible for autonomous vehicles to wait for the generation completion of \acp{llm} when taking crucial decision or transmitting important information to surrounding vehicles. Inference acceleration techniques and architectures on both system level and algorithm level \cite{Wan2023EfficientLLM} e.g., KV caching \cite{zhang2024h2o}, FlashAttention \cite{dao2022flashattention} and \acfp{moe} \cite{shazeer2017outrageously} could largely increase the throughput of \acp{llm} (tokens per second) to alleviate this issue. Finally, even physical challenges such as insufficient memory and low high latency are mitigated by the combination of various techniques, it remains a fundamental difficulty for \acp{llm} to accomplish telecom-specific tasks in wireless networks due to a general lack of knowledge in telecom domain. Therefore, it would be natural to anticipate the existence of telecom-specialized or telecom-specfic \acp{llm}, which is exactly the core problem this paper tries to tackle with. Before diving into the technical details of our proposed methods, we briefly review the recent advances including domain-specific \acp{llm}, applications of \acp{llm} in telecom and the the challenge of building telecom-specific \acp{llm}.

\subsection{Related Works}

\textbf{Domain-Specific \acp{llm}}: Although general-purpose \acp{llm} demonstrate considerable capabilities in various tasks, their performance degenerates seriously for tasks which necessitates domain-specific knowledge, e.g., math, finance and Telecom. Therefore, domain-specification will be mandatory to enhance \acp{llm}'s performance in these domains. For instance, BloombergGPT \cite{wu2023bloomberggpt} is the first financial \acp{llm} which is pre-trained on massive mixed data of financial and general corpora. Due to the high cost of pre-training \acp{llm} from the scratch, FinGPT was proposed by fine-tuning general-purpose \acp{llm} on 34 online curated data sources and through \acf{rag} \cite{yang2023fingpt, zhang2023instruct, wang2023fingpt, liu2023fingpt, zhang2023enhancing}. To enhance math problem solving capabilities of \acp{llm}, \acf{rleif} is proposed in \cite{luo2023wizardmath} to train WizardMath which outperforms various general purpose \acp{llm} and math-specialized \acp{llm} in GSM8k \cite{cobbe2021training} and MATH \cite{hendrycks2021measuring}.  Other examples of domain-specific \acp{llm} include MedPalM2 in medical domian, ChatLaw \cite{cui2023chatlaw} and SaulLM \cite{colombo2024saullm} in legal domain, and Code Llama \cite{Rozire2023CodeLlama} in code generations. For more details of domain-specific \acp{llm} such as training dataset and training methods, see examples in Table \ref{tab:domain_specific_LLM}.

\begin{table*}
\centering
\resizebox{\textwidth}{!}{%
\begin{tabular}{ | c | c | c | L | } 
\hline
\acp{llm} & Domain & Data & Training Method \\
\hline \hline
MedPaLM2  \cite{singhal2023towards} & Medical & Multiple Med QnA Datasets  & \acf{sft} \\ 
\hline
ChemLLM  \cite{zhang2024chemllm} & Chemistry & Scientific Publications & Two-stage \ac{sft} \\ 
\hline
BloombergGPT \cite{puri2019zero} & Finance & Financial + General & Pre-training \\ 
\hline
FinGPT \cite{yang2023fingpt, zhang2023instruct, wang2023fingpt, liu2023fingpt, zhang2023enhancing, zhang2023fingptrag} & Finance & Financial & \ac{sft}+ \ac{rag} \\ 
\hline
Xuanyuan \cite{zhang2023xuanyuan} & Finance & Finance + General & Continue Pretrain + Hybird Tuning \\ 
\hline
BioGPT \cite{luo2022biogpt} & Biology & PubMed & Pre-training \\ 
\hline
BioMistral \cite{labrak2024biomistral} & Biology & PMC Open Access Subset & Continue Pretrain + \ac{sft} \\ 
\hline
ChatLaw \cite{cui2023chatlaw}  & Legal  & Chinese Legal Documents & \ac{sft} + \ac{rag} \\ 
\hline
SaulLM \cite{colombo2024saullm}  & Legal  & Legal Documents & Continue Pretrain + \ac{sft} \\ 
\hline
Galactica \cite{taylor2022galactica} & Science  &  Scientific Documents & Prompt Pre-training \\ 
\hline
Mozi \cite{ni2024mozip} & Science  &  RedPajama \cite{together2023redpajama} Arxiv & \ac{sft} + RAG \\ 
\hline
Llema \cite{azerbayev2023llemma} & Math  & Proof-Pile-2 & Continue Pretrain  \\ 
\hline
WizardMath \cite{luo2023wizardmath} & Math & GSM8k \cite{cobbe2021training} + MATH \cite{hendrycks2021measuring} & \ac{sft} + RLEIF \cite{luo2023wizardmath} \\
\hline
DeepSeek Math \cite{shao2024deepseekmath} & Math & DeepSeek Math Corpus & Continue Pretrain + \ac{sft} + RLHF \\
\hline
StarCoder 2 \cite{Lozhkov2024StarCoder2} & Code & The Stack 2 \cite{Lozhkov2024StarCoder2} & \acf{fim} \cite{bavarian2022efficient} \\ 
\hline
Code Llama \cite{Rozire2023CodeLlama} & Code & Proprietary & Continue Pretrain + FIM + SFT \\ 
\hline
Codestral \cite{Codestral}  & Code & Proprietary & Pre-training  \\ 
\hline \hline
\textit{TelecomGPT} & \textit{Telecom} & \textit{Public Telecom Documents} & \textit{Continue Pretrain + SFT + DPO \cite{rafailov2024direct}} \\  
\hline\hline
\end{tabular}}
\caption{Domain Specific LLMs: domains, data or datasets and training methods.} \label{tab:domain_specific_LLM}
\end{table*}

\textbf{\acp{llm} in Telecom}:
\acp{llm} has gathered a lot of attentions in telecom domain recently. It is not surprising that \acp{llm} can perform  \acf{nlp} tasks such as fault analysis \cite{chen2023tele} and \ac{tdoc} classifications \cite{bariah2023understanding} within the context of telecom. In \cite{tarkoma2023ai}, it is shown that two distinct nodes interact seamlessly for enhanced network optimization, demonstrating the potential of \acp{llm} in varied telecom-relevant scenarios. Other examples of applying general purposed \acp{llm} to use cases such as generating network configuration \cite{wang2023making,mondal2023llms} and assisting FPGA-based wireless hardware development \cite{du2023power} showcase the versatility of \acp{llm} in tackling domain-specific tasks within the telecommunications sector. Apart from existing applications, \acp{llm} are envisioned to be key enabling techniques such as semantic communications \cite{zou2023wireless,park2023towards,jiang2023large}, intent-driven networks and self-evolving networks \cite{bariah2024nextbigthing}. In \cite{Lin2023PushingLL}, \acp{llm} can be effectively deployed in end user and vicinity thorough edge training and edge inference. 
\ac{rag} is used in \cite{Xu2024LargeMM} to align the math and reasoning capabilities of multi-modal \acp{llm} to wireless system design. A semantic-native \ac{llm}-based network is proposed in \cite{Zou2024GenAINet}, where Telecom agents use \acp{llm} to extract semantic information from raw data (observations), learn reasoning path towards decisions (responses), distillate knowledge from memory and exchange in a network to reduce the overall energy consumption of the system. These early explorations demonstrate an obvious paradigm shift compared to conventional \acf{dnn} approaches.


\textbf{Challenges of building Telecom-specific \acp{llm}}: Despite the promising benefits of deploying \acp{llm} in a  centralized or decentralized manner in different networks, mainstream \acp{llm} lack seriously knowledge and know how of telecom domain. Practical telecom networks operates in protocols and standards in pre-defined in \acfp{sdo} such as \acf{3gpp}, \acf{ieee} and \acf{itu}. However, \ac{sota} models such as GPT-4 fails almost half part of the specification-related problem in TeleQnA \cite{Maatouk2023TeleQnAAB}, hindering the potential deployment in networks. Therefore, it is crucial to enhance current \acp{llm} telecom knowledge in all aspects to be integrated into current networks seamlessly. To achive this taraget, there are (not limited to) several main challenges to build a Telecom-specific \ac{llm}:
\begin{itemize}
\item Missing of Telecom text datasets: despite various amount of textual documents available in telecom domain, there are almost no open-source dataset available for pre-training or fine-tuning a telecom-specific \ac{llm}.
\item Technical documents dominance: \acp{tdoc}, specifications and research papers written in a professional manner dominates the overall text files, making the training of \acp{llm} challenging.
\item Multi-modality contents: important concepts, scenarios and methods are usually explained and represented in different formats, e.g., figures, tables in standards and patents.
\item Frequent knowledge updating: regular release announcement from \ac{sdo} and the rapid advancement of research publications make it difficult for \acp{llm}, with their high training cost and long training time, to be updated in a  timely manner.
\end{itemize}
Therefore, instead of pre-training a Telecom \ac{llm} from scratch which is expensive and unprepared for the time being, it would be efficient and reasonable to consider adapting general-purpose \acp{llm} to telecom domain under acceptable cost and training time, which is exactly the target of this paper.

\subsection{Contributions}

Our main contributions  are summarized as follows:
\begin{itemize}
\item We collect and build a telecom-specific pre-training dataset to build a telecom-specific \acp{llm}.
\item We construct a high quality telecom-specific instruction dataset with instructions for diverse tasks in telecom domain.
\item We propose effective benchmarks to evaluate crucial capabilities of \acp{llm} such as such as knowledge query, math modeling, \acp{tdoc} classification, code generation and analysis in telecom context.
\item We propose a pipeline to adapt a general purpose \acp{llm} to a telecom-specific \ac{llm} which consists of continual pretraining, instruction tuning and alignment tuning within the context of telecom.
\item We train a telecom-specific \ac{llm} which can be used for multiple downstream tasks in telecom domain.
\end{itemize}
The remaining of the paper is organized as follows. Firstly the methodology of adapting a general purpose \ac{llm} to a telecom-specific \ac{llm} is detailed in Sec. \ref{sec:methodology}. In Sec. \ref{sec:dataset}, we present our pipeline of building datasets for different stages of training.  After that, we present the evaluation benchmarks to evaluate the key capabilities of an \ac{llm} for telecom-relevant tasks in Sec. \ref{sec:benchmark}. Then, our training setting and the main results obtained are in Sec. \ref{sec:training} and Sec. \ref{sec:results}. Finally, we conclude the paper in Sec. \ref{sec:conclusion}.
\section{Methodology}
\label{sec:methodology}

Our method of building telecom-specific \ac{llm} consists of three standard stages, namely, domain specific continual pre-training, instruction tuning and alignment tuning. 

\subsection{Telecom-Specific Continual Pre-training}
Different from pre-training from the scratch for domain-specific \acp{llm} such as BloombergGPT for finance and Galactica \cite{taylor2022galactica} for scientific domain, continual pre-training \cite{gururangan2020don,ke2023continual} offers an alternative approach in contrast which requires considerably less cost to specialize general-purpose \acp{llm}, e.g., Code Llama \cite{Rozire2023CodeLlama} Deepseek Math \cite{shao2024deepseekmath} and SaulLM \cite{colombo2024saullm}. Similar to pre-training stage, the training objective is the causal language modeling, i.e., predicting the next token conditioned on previous tokens. Denote $x = \left(x_1, \dots, x_T \right)$ a sequence of tokens representing a text input and $\theta$ the parameter of an \ac{llm}, the causal language modeling task is to minimize the negative log-likelihood loss expressed as:
\begin{equation}
    \mathcal{L} \left(x,\theta \right) = - \sum_{t=1}^T \log  \mathds{P} \left(x_t \middle| x_{<t}\right),
\end{equation}
where $x_{<t}$ denotes the token sequence before token $x_t$.

\subsection{Telecom-Specific Instruct Tuning}
During (domain-specific) pre-training, \acp{llm} learn and acquire general knowledge but might not not excel in user interactions. Therefore, pre-trained \acp{llm} should be further instruct tuned to follow instructions to interact with users and other \acp{llm} through \acf{sft}. With instruct tuning, \acp{llm} are capable to follow instructions even for unseen tasks without explicit examples. Instruct tuned \acp{llm} generally have better performance in zero-shot or few-shot scenarios and fewer refusals when responding to users' requests. For a given instruction dataset consisting of multiple instruction-response pairs: $\mathrm{I} = \left \{ x^{(i)}, y^{(i)} \right \}_{i = 1}^N$ with instruction $x^{(i)}$ and corresponding response $y^{(i)}$, the training objective is to minimize the negative log-likelihood of a response conditioned on its instruction:
\begin{equation}
    \mathcal{L} \left(y^{(i)},\theta \right) = - \sum^{|y^{(i)}|}_{t=1} \log  \mathds{P} \left(y^{(i)}_t \middle| y^{(i)}_{<t} , x^{(i)}\right)
\end{equation}

\subsection{Telecom-Specific Alignment Tuning}

After supervised fine-tuning of  pre-trained \acp{llm}, then we need to further align the model with human preference. \acf{rlhf} is proposed in InstructGPT to improve the helpfulness of \acp{llm} \cite{Ouyang2022TrainingLM}. However, RLHF requires preference data collected from human labeler and building reward model which is both costly and unstable during the reinforcement learning due to the existence of \acf{kl} divergence in the training objective and discrete nature of language generation. Instead, \acf{dpo} \cite{rafailov2024direct} is proposed to mitigate the difficulty in \ac{rlhf}. Instead of constructing an explicit reward function, a Bradley-Terry model \cite{Bradley1952RankAO} is applied to reparameterize the reward so that the probability of preference of the ground truth and the generated answer is independent of the reward model, significantly simplying the training procedure. For \ac{dpo}, only a simple binary preference dataset consisting of pairs of chosen and rejected samples, is required.  The objective of \ac{dpo} is formulated as:
\begin{align}
&\mathcal{L}_{\text{DPO}} (\pi_{\theta}; \pi_{\text{ref}}) \notag \\
= & -\mathbb{E}_{(x, y_w, y_l)} \Bigg[ \log \sigma \Bigg( \beta \log \frac{\pi_{\theta}(y_w | x)}{\pi_{\text{ref}}(y_w | x)} - \beta \log \frac{\pi_{\theta}(y_l | x)}{\pi_{\text{ref}}(y_l | x)} \Bigg) \Bigg],
\end{align}
where $(x, y_w, y_l)$ denotes a preference pair, with $x$ being the prompt. In this pair, $y_w$ is the response preferred by the human labeler or LLM judge, and $y_l$ is the less preferred response; $\beta$ is a parameter controlling the deviation from the base reference policy $\pi_\text{ref}$;  $\sigma \left( \cdot \right)$ stands for the logistic function.  We select the reference model $\pi_\text{ref}$ as the \ac{sft} model we obtained in previous stage.

\begin{figure*}[t!]
\centering
\includegraphics[width=1\linewidth]{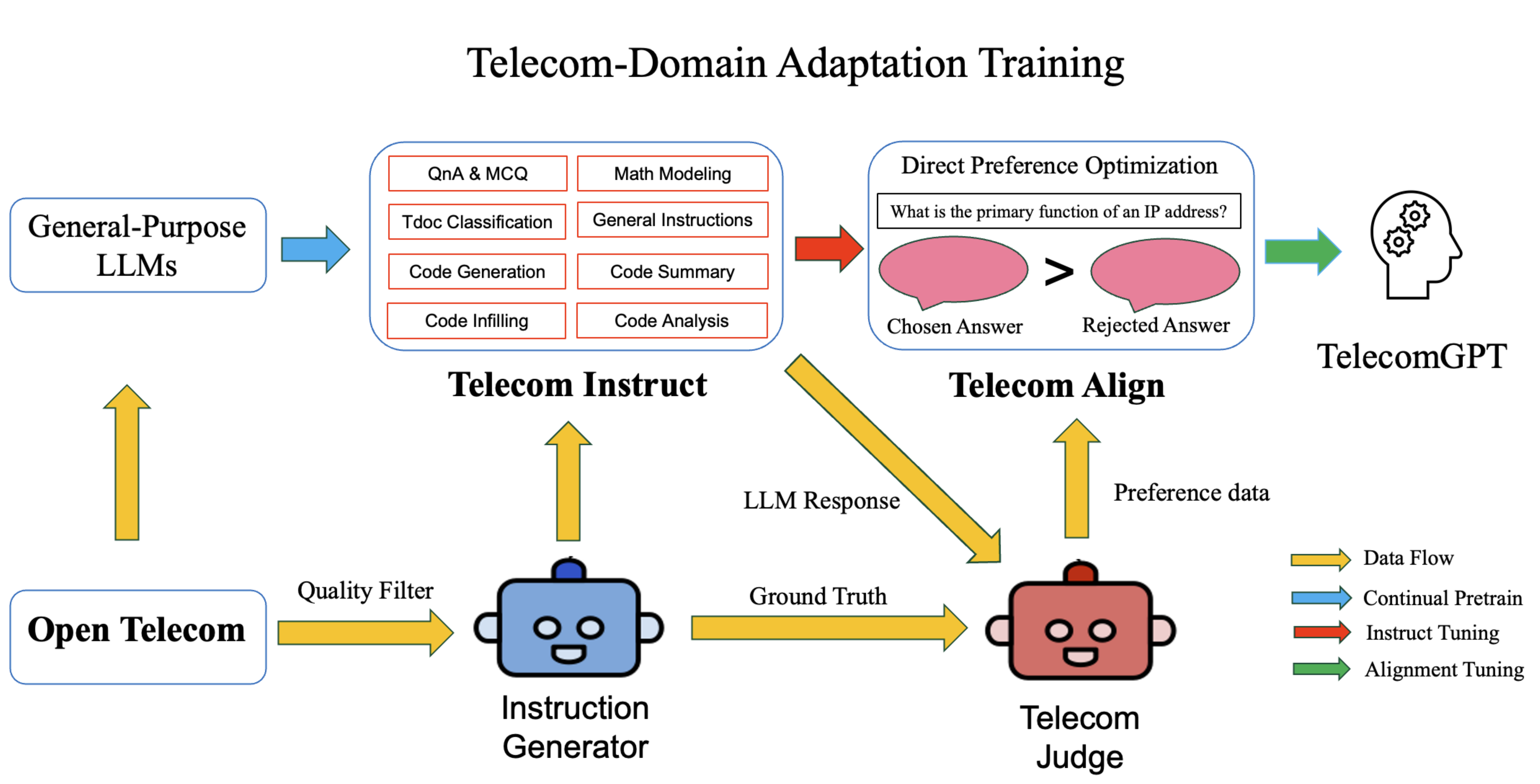}
\caption{The training pipeline of our TelecomGPT framework. The full pipeline consist of three training stage, namely, continual pretraining on telecom domain, instruct tuning (\ac{sft}) and alignment tuning.}
\label{fig:training_pipeline}
\end{figure*}

\section{Dataset}
\label{sec:dataset}
Following our training method, we need to build three datasets, namely the pre-training dataset for continual pre-training, the instruction dataset for instruction tuning and the preference dataset for alignment tuning.

\subsection{Pre-training Dataset - \textbf{OpenTelecom}}

The pre-training dataset is collected mainly from web, which includes Telecom standards, research papers, books, patents, \acp{qa}, wiki, and codes, as detailed in Table \ref{tab:OpenTelecomData}. We developed the following two preprocessing pipelines according to the source of these data.

\textbf{Telecom Standards}:
\ac{3gpp} is the main \ac{sdo} in the area of Telecommunication. The universal standards for 3G, 4G and 5G have been developed by \ac{3gpp} since 1999. \ac{3gpp} works with \ac{tdoc} contributed by companies during the development phase and produces technical specifications as a final output. The specification work is carried out in Technical Specification Groups (TSGs). There are three Technical Specifications Groups: \ac{ran}, \ac{sa}, and \ac{ct}. Each TSG consists of multiple \acp{wg} focused on specific areas, ranging from radio access network specifications, core network specifications, service requirements and specifications, and architecture and protocols for mobile communication systems, to \ac{qos} and performance requirements, security and privacy in mobile communication systems, interoperability and compatibility requirements, network management and operation, and testing and certification procedures. We scrap the technical specifications, reports, and documents from release 8 to 19 on 3GPP FTP site. IEEE is another important \ac{sdo} developing popular Telecom standards, such as WiFi and Ethernet. We scrap the standard documents of IEEE 802.11, 802.3, 802.15, and C95.1 series from IEEE Get program. 

We utilize similar methods in \cite{bariah2023understanding} to preprocess the raw standard documents, including: 1) Parse the HTML tags in the text and return the text content without any HTML tags using \textit{BeautifulSoup}; 2) Remove any URLs (web links) from the text: identify the regex pattern that matches URLs starting with either "http" or "https" and may include alphanumeric characters, special characters, and encoded characters. 3) Remove tables from the parsed HTML document using \textit{BeautifulSoup}. 4) Divide each document into multiple text segments with different number of words extracted from natural language toolkit (NLTK). This allows us to evaluate the model's capability of understanding technical descriptions in different lengths. 5) Remove headers, footers, captions, and pseudo codes. Also, we eliminate the references section and all the text afterward. 6) Remove \acp{cr}, drafts, templates due to their limited technical information.  

\textbf{Telecom Content Filtering from General Dataset}: Despite the fact that no open-sourced telecom-specific pre-training dataset exists, there are gigantic amount of telecom content in general purpose pre-training dataset such as C4 \cite{raffel2020c4}, RefinedWeb \cite{Penedo2023TheRD} and Redpajama \cite{together2023redpajama}.  We select Rejama-1T to demonstrate how to filter telecom-relevant content from general dataset to take advantage of its categorical structure and also due to our hardware limits. RedPajama-1T is an open pre-training dataset consisting 7 subsets: Commoncrawl, C4, Github, Books, Arxiv, Wikipedia and StackExchange. The entire dataset contains $1.2$ billion text tokens. In order to filter all telecom-relevant contents from open dataset, there are two steps in our framework: keyword filtering and deduplication.


To select Telecom-specific data from a general-purpose dataset, we collect and select a non-exhaustive $700$ keywords from telecom textbook, section of definitions and abbreviations from open 3GPP technical specifications and etc. We propose 6 criteria for selecting a keyword so that it can effectively help us to filter telecom-specific content:
\begin{itemize}
\item  Domain Specificity: Keywords are closely associated with the telecommunications industry and theory. Positive Examples: ``5G", ``VoIP", ``MIMO" and ``LTE Advanced"; Negative Examples: ``Network", ``Service" and ``Platform".
\item Frequency in Telecom Discourse: Keywords frequently appearing in telecom discussions and publications. Positive Examples: ``Broadband", ``LTE", ``Roaming" and ``Signal strength"; Negative Examples: ``Connection", ``Speed" and ``Improvement".
\item Distinctiveness within Telecom: Keywords uniquely identifying telecom technologies or services. Positive Examples: ``Spectrum allocation", ``Fiber optic", ``Beamforming", ``Cellular handoff"; Negative Examples: ``Data transmission", ``Wireless", ``Communication system"
\item Authority and Standards: Keywords from telecom standards and regulatory documents. Positive Examples: ``3GPP", ``IEEE 802.11", ``ITU-T", ``E.164"; Negative Examples: ``Software update", ``User interface design"
\item Timeliness and Emerging Technologies: Keywords reflecting the latest advancements in telecom. Positive Examples: ``Network slicing", ``Semantic communication", ``Quantum cryptography", "5G NR (New Radio)"; Negative Examples: Technologies no longer at the forefront like ``2G", ``PSTN", ``ISDN".
\item  Clarity and Avoidance of Ambiguity: Avoid vague or broadly used terms. Positive Examples: ``VoLTE", ``Wi-Fi 6", ``SD-WAN", ``IPv6 adoption"; Negative Examples: Broad or ambiguous terms like ``Transformation", ``Efficiency", ``Fourier Transform", ``Convex Optimization".
\end{itemize}

Most keywords in our keyword list satisfy more than 4 criteria out of 6. Both the keywords and their abbreviations is included in the keywords list. However, to avoid introducing irrelevant content due to the potential polysemy of abbreviations, we test our keyword list on a small portion of the entire dataset, and remove those abbreviations frequently introducing noise, e.g., DL (deep learning v.s. downlink) and SAP (Service Access Point v.s. Systems Applications and Products). For each data sample, we introduce two quantities to evaluate its relevance to telecom domain: number of telecom keywords (same keyword will only be counted once) and telecom keyword density. The density for a text of $N$ words with a total number of $M$ telecom keywords matches is $ M /\log(N+1)$. The reason for using the logarithm of the number of total words rather itself is to compensate those long documents. For example, a long text of 1000 words with 30 telecom keywords would be considered to be less relevant to telecom compare to a short text of only 100 words with only 1 telecom keyword which is an absurd. Besides, the number of total keywords is also an effective metric to avoid obtain false positive data samples in the final dataset.

Deduplication refers to the process of removing redundant or repeated contents in a dataset. Deduplication is crucial for improving efficiency of pre-training \cite{lee2021deduplicating} and for mitigating diversity decreasing \cite{hernandez2022scaling}.
For telecom-specific \acp{llm}, this process is unavoidable since most standards or protocols are generally approved by multiple \acfp{sdo}, e.g., technical specification of \ac{3gpp} and \ac{esti}. Moreover, multiple releases of some topics introduces redundant contents naturally. For the sake of simplicity, we directly apply existing deduplication tool in xorbits \cite{xorbits}.

\begin{table*}[t!]
\centering
\begin{tabular}{ | L | c | c | c | } 
\hline \hline
Category & Training & Validation & Percentage (\%) \\
\hline \hline
3GPP Standard & 193 & 1.9 & 11.49 \\ 
\hline
IEEE Standard & 7.5 & 0.07 & 0.45 \\ 
\hline
Paper (arxiv) & 893  & 9 & 53.17 \\ 
\hline
Books & 1.9 & 0.02 & 0.11 \\ 
\hline
Patent (C4) & 253.2 & 2.6 & 15.08 \\ 
\hline
StackExchange & 51.9 & 0.5 & 3.09 \\ 
\hline
Wikipedia & 18.9 & 0.2 & 1.13 \\ 
\hline
Code (Github) & 260.1 & 2.6 & 15.49 \\ 
\hline
Total & 1679.5 & 16.89 & --- \\ 
\hline
\end{tabular}
\caption{OpenTelecom dataset for Continual Pre-training (in million tokens)} \label{tab:OpenTelecomData}
\end{table*}

\subsection{Instruct Tuning Dataset - \textbf{TelecomInstruct}}

To build an effective instruction dataset, it is widely shown that the diversity of different tasks in the dataset is more important than the number of instructions of same type. Our instruction dataset, referred as  \textbf{Telecom Instruct} in what follows consists of different critical tasks in telecom domain:
\begin{itemize}
\item  \textbf{\ac{mcq} Answering}: Select all correct answers from a \ac{mcq}.
\item  \textbf{Open-ended Question Answering}: Answer telecom-relevant question from standards, research papers or patents in an open-ended manner.

\item  \textbf{Technical documents classification}: Classify text from different Tdocs into the related working group as documented in the library of \acp{sdo}.

\item  \textbf{Math Modeling}: Generate accurate math equation such as channel models for given text description of system model and problem formulation in the context of telecom with predefined notations.

\item  \textbf{Code Generation}: Generate script/function for a given task/functionality in telecom domain such as sending signal indicator and extracting MAC address from a frame.

\item \textbf{Code Infilling}: Infill incomplete script based on the context and the targeted functionality. This task is also known as \acf{fim} \cite{bavarian2022efficient} task which is beneficial for developers or researchers to improve efficiency when generating telecom-relevant scripts.
\item  \textbf{Code Summary}: Summarize the core functionality of a given script, including identifying if the script is telecom-relevant or not.
\item \textbf{Code Analysis}: Detail the operational logic behind the function, emphasizing the knowledge and principles in telecom domain.
\item \textbf{General Instruction}: Explain concepts, describe specifications, identify problems, propose solutions, summarize content in Telecom standards, patents, and papers.
\item \textbf{Protocol Instruction}: Generate the protocol workflows in Telecom standard following a human prompt. 

\end{itemize}

Instructions such as \ac{tdoc} classification, math modeling and code infilling can be easily built for raw text, papers and codes. For the remaining tasks, we utilize large \ac{sota} models, i.e., GPT3.5 or GPT4 to generate instructions with well-designed prompts, plus expert filtering and validation. The details of how these instructions are generated will be demonstrated and explained in Sec. \ref{sec:benchmark} for following the same methodology as building corresponding benchmarks.

\subsection{Alignment Tuning Dataset -  \textbf{TelecomAlign}}
After \ac{sft} on our telecom instruction dataset, we found that the SFT models have learned how to perform telecom-relevant tasks in general. However, we can still observe undesired response such as repeated generation, too short response and telecom-irrelevant content generation. Rather than collecting real human preference data which is costly and inefficient, we simply define our preference as \emph{\textbf{to provide concise and accurate answer with minimum amount of information unless requested especially}}. Such preference is reasonable to reduce the latency introduced in a \ac{llm}-based communication system and aligns with concepts such as semantic communication in a naive sense. Other response preferences are also possible. Here we just demonstrate how to align \acp{llm}'s response with human preference for telecom-relevant tasks. Besides, such preference can be easily measured by comparing metrics like Rouge scores and length of the ground truth and \acp{llm}' response. Therefore, our preference dataset is obtained by selecting instructions with low performance metrics. Pre-trained \acp{llm} can be utilized to select preferred response when provided with clear preference description. 




\section{Evaluation Benchmarks}

\label{sec:benchmark}

After continual pre-training, instruct and alignment tuning, a general purpose \acp{llm} is adapted to telecom domain. We would like to evaluate the performance of obtained telecom-specific \ac{llm} in some critical tasks in telecom domain including both the academic and industrial applications. Unfortunately, there are few telecom-specific evaluation benchmarks available, e.g., TeleQnA \cite{Maatouk2023TeleQnAAB}. Moreover, we believe that an \ac{llm} with extremely high accuracy in \ac{mcq} alone might not meet the requirements of practical applications such as the tasks we mentioned in the instruct tuning phase. Therefore, it would be mandatory to construct reasonable benchmarks to evaluate the capabilities of a telecom-specific \ac{llm} in various downstream tasks.

\subsection{Telecom Question Answering}

\textbf{Multiple-Choice \ac{qa}}: Based on the pipeline of creating TeleQnA dataset, we utilize GPT-4 to generate the questions. We first clean the raw telecom-relevant texts such as ArXiv paper, Wikipedia pages, patents and textbooks by removing the references in case that the generated questions are related to any specific literature. After that we use two \ac{llm} agents to automate the generation process. The cleaned text is given to the first GPT-4 agent with the prompt illustrated in Fig. \ref{fig2:mcq_template}. 
After that, we feed the original text and the generated question into the second GPT-3.5 agent to validate the correctness of the generated question. If a correct answer is obtained, we will keep the question-answer pair, otherwise it is removed. Finally, we conduct human validation to ensure the correctness of generated question-answer pairs. 

\textbf{Open-ended \ac{qa}}: We utilize the same method to create an evaluation dataset of non-contextual \ac{qa}, where all options are removed. In this benchmark, an \ac{llm} generates open-ended answer to a question, which is evaluated over the ground-truth answer using Rouge scores. It makes the evaluation close to the real \ac{qa} use cases. The dataset is also used to align \ac{llm} with Telecom preferred responses during instruct and alignment tuning.

\begin{figure}
\begin{mdframed}[linecolor=black, linewidth=2pt, roundcorner=10pt]
\textbf{Prompt Template for Telecom Question Answering}
\hrule height 0.5pt
\vspace{0.15in}
\textbf{Instruction:} Please create several multiple-choice questions based on the provided texts. These created questions must be generated in this form:
\begin{itemize}
    \item {Question:} XXXX
    \item {Option 1:} XXXX
    \item {Option 2:} XXXX
    \item ...
    \item {Answer:} Option X
    \item {Explanation:} explain why the correct answer is Option X.
\end{itemize}

\vspace{0.1in}
These questions should not refer to any equation. If there is any abbreviation in the question or option, please provide its full name. The ``Answer" must be in the format of ``Option X". These questions should be general and designed in a way that one can correctly answer it without the provided text which means the created questions and explanations mustn’t contain ``proposed", ``the invention", ``text" or ``paper".

\vspace{0.1in}
\textbf{TEXT}:
$\left \{ \text{text} \right\}$
\end{mdframed}
\caption{Prompt template for creating \acp{mcq}. Clear instructions are given to \acp{llm} to avoid references.}
\label{fig2:mcq_template}
\end{figure}

\subsection{Telecom Standard Documents Classification}
One key capability of a telecom-specific \ac{llm} is to understand and manage technical documents. There are three aspects why such knowledge is vital.  Firstly, when envisaging an \ac{llm} (agent) acting as a decision maker for the network management, it is mandatory for an \ac{llm} to understand understand the specific scenario it is facing within the established standards and protocols. Second, the knowledge of the valid range of an action is essential. For example, the uplink transmit power control (TPC) command filed with reference to certain accumulated and absolute power tuning as defined in 3GPP TS38.213. 
Lastly, the knowledge of standards and protocols allows an \ac{llm} to predict the impact of its action and the reaction of other entities. In summary it is difficult for any \acp{llm} to operate in practical networks if a deep and accurate understanding of the technical documents is missing. To evaluate these capabilities, a simple but efficient way to evaluate the basic knowledge of \acp{llm} is to first identify the relevance of any text to a given \ac{tdoc}. Therefore we follow the same methodology in our previous work \cite{bariah2023understanding} to classify a random text extracted from a \ac{tdoc} into one unique working group of \ac{3gpp}: CT1, CT3, CT4, CT6, RAN1, RAN2, RAN3, RAN4, RAN5, SA1, SA2, SA3, SA4, SA5 and SA6. No context information will be further provided to \acp{llm}. Despite the extreme simplicity of this task, as we will show later in numerical results, it is difficult for most mainstream general purpose \acp{llm} to predict accurately working group in zero-shot setting.



\subsection{Telecom Math Modeling}  
To undertake a telecom-related task, a critical capability for a telecom-specific \ac{llm} is to accurately model the pertinent environment or problem using telecom terminology and to formulate the task as rigorously defined problems expressed through precise mathematical equations. Requesting an \ac{llm} to perform system modeling and problem formulation requires both in-depth telecom knowledge and factual reasoning capability. Therefore, we believe a benchmark capable of evaluating the math modeling capabilities of \acp{llm} within the context of telecom provides an indispensable perspective to a comprehensive evaluation of a telecom-specific \ac{llm}. Nevertheless, due to the flexibility of modeling (considering the myriad potential assumptions and extensive variety of existing models in the literature), it is challenging to directly evaluate the effectiveness of a system model or the problem formulation generated by \acp{llm}. Hence, building a simple-but-effective benchmark is crucial to evaluate the problem modeling and formulation capabilities of an \ac{llm}. 

Therefore, we propose the following masked equations infilling task for \acp{llm}: for a research paper or any technical document, we first extract those paragraphs relevant to the system modeling and problem formulation. 
Then we mask some crucial equations for the modeling and problem formulation process, e.g., system model, channel model, coding and decoding schemes. This type of masked language modeling task is widely used for language understanding task, e.g., in BERT \cite{devlin2018bert} and its variants and has been proven to be useful while equations are masked entirely here.  A masked equation will be replaced by a special placeholder $<\text{MASK}>$ to indicate that there is a missing equation.  In the meantime, due to the uncountable possibilities of different notations, we try to avoid mask those equations related to definitions, referring other documents (typically defined inline). Finally, an \ac{llm} will be asked to recover these equations and the predicted equation would be compared to the ground-truth to obtain a score based on the quality of prediction. 
To further simplify the task, an LLM is merely asked to predict one equation at a time. The context, along with the ground-truth equations (rather than the predicted equations from previous rounds), is provided before the equation to be predicted.

\begin{figure}[t!]
\begin{mdframed}[linecolor=black, linewidth=2pt, roundcorner=10pt]
\textbf{Prompt Template for Telecom Code Generation}
\hrule height 0.5pt
\vspace{0.15in}
\textbf{Instruction:} As a distinguished expert in telecommunications software, you are skilled in developing and optimizing programs within the telecom domain, using languages like C++, C, Python, and Matlab. When provided with a script or function, your task is to craft a concise yet comprehensive request that can guide the generation of the script, if it is telecom-relevant. Refer to the examples provided below and tailor your request accordingly:

\begin{itemize}
    \item \textbf{Example 1:} Write a Python function to convert an IPv6 address from string format to an integer.
    \item  \textbf{Example 2:} Develop a C function that updates the decrypt status flag based on the decryption result for a received 802.11 frame.
\end{itemize}

Your request should satisfy the following requirements:
\begin{itemize}
    \item Clearly identify the programming language of the script; 
    \item Describe the script's functionality in telecom domain.
    \item If necessary, describe the scenario or environment where the script operates.
\end{itemize}

\vspace{0.1in}

Should the script is telecom-relevant, please respond with ``response": ``content of your request". If the script lacks strong relevance to telecommunications or seems to be an incomplete function, your response should be ``response": ``irrelevant". Please ensure your request directly pertains to the script's utility in the telecom sector, omitting any unnecessary details or commentary.

\vspace{0.1in}
\textbf{SCRIPT}:
$\left \{ \text{script} \right\}$

\end{mdframed}
\caption{Prompt template for creating code generation instruction. Clear instructions and examples are given to \acp{llm} to filter telecom-irrelevant functions or avoid generating meaningless request.}
\label{fig:code_generation_template}
\end{figure}

It remains how to evaluate a predicted equation to its ground truth. We utilize MathBERT \cite{shen2021mathbert}, a variant of BERT fine-tuned on a large dataset of math equations, to evaluate the average cosine similarity between the embeddings of the predicted equation and the ground truth. The cosine similarity of MathBERT is adapted to the operation trees of the equations and thus can be used as a semantic similarity in the sense of math structure.  For a given equation represented by $y$ and a predicted equation $\widehat{y}$ given by an \ac{llm}, we define the MathBERT score as:
\begin{align}
 & \text{score} \left(y,\widehat{y} \right) \nonumber \\
 = & \max \left \{ \frac{\cos \left(e\left(y\right),e\left(\widehat{y}\right) \right) - \cos \left(e\left(y\right),e\left(\varnothing\right) \right)}{1 - \cos \left(e\left(y\right),e\left(\varnothing\right) \right)}, 0 \right\} \times 100\%
\end{align}
where $e\left( \cdot\right)$ is the embedding output of the MathBERT; $\cos$ denotes the cosine similarity between two embedding vectors; $\varnothing$ represents an empty equation, i.e., an \ac{llm} returns nothing. The MathBERT is actually a normalized cosine similarity of range $\left [0,100\% \right]$  w.r.t. the difference between the ground truth and the ``empty" answer. This normalization is introduced since the cosine similarity between the ground truth and an empty answer is usually greater than $0.7$. During our tests, a random generated equation can easily achieve a cosine similarity greater than $0.8$. Therefore, we believe the raw cosine similarity metric of MathBERT is not aligned with common sense of human. See appendices for example demonstrations with fundamental equations in telecom domain.

\subsection{Telecom Code Understanding and Generation}
Practical telecom systems operate in codes written in different languages. To understand and generate codes for a well-defined telecom task given in standards or protocols are vital for telecom-specific \acp{llm}. Besides, code analysis and generation capabilities are important for an \ac{llm} to interact with existing simulators. To start with, we consider four common programming languages used in telecom domain: C, C++, Python and Matlab. Our Code evaluation benchmark include four tasks mentioned in our instruction dataset. 

Other standard code-relevant task, e.g., code fixing/editing is not included for lack of relevant open-sourced dataset. The data samples used to build our benchmark are selected following the same logic in the telecom content filtering process for pre-training dataset. We further select those code files with larger number of telecom keywords, number of copies and keyword density to ensure the quality of data and the relevance to telecom domain. Then we extract telecom-relevant function blocks from these files.  For code generation, we prompt GPT-4o to generate corresponding generation requests using the prompt illustrated in Fig. \ref{fig:code_generation_template}. The summary and analysis instructions can be obtained using similar prompts and we omit them here to avoid repetitions.  For code infilling task, several successive lines in the middle of a code file are randomly replaced with a special placeholder $<\text{FILL}>$, the missing part of code is considered to be ground truth. Due to the complexity of telecom-relevant script, it is difficult to directly evaluate the effectiveness of the script similar to common benchmark such as HumanEval \cite{chen2021humaneval}. Instead, we utilize Rouge score to compare the similarity between generated script and the ground truth.  Although being hard to judge the quality of generating code lines for code generation and infilling tasks, it provides a reasonable metric on the relevance of generating scripts in telecom domain rather the true effectiveness. 

\subsection{Telecom Instruct Following}

Using \acp{llm} to improve the productivity of Telecom system research, design, and development is a key application in Telecom industry. For example, it can help researchers to find solutions in the literature for a Telecom problem, or guide engineers to identify the key steps to implement a Telecom feature in the standard. In order to enhance TelecomGPT's instruct following capabilities preferred by Telecom users, we further develop the following two benchmarks: 
\begin{itemize}
    \item \textbf{General Instruction}: We utilize GPT models to create a number of instruction tasks from standard documents, specifications, research papers, and patents. Example tasks include identify issues in a specified scenario, explain a technical feature in standard, find possible solutions of a research problem, and so on. The prompt used to generate the data comprises: 1) Description of the task, formats and rules to avoid referring to specific figures, tables, sections, etc.; 2) Some examples of the instruction tasks; 3) Text segments from the pre-train dataset. The output includes \textit{instruction} of the task, \textit{input} of context information such as scenario description, and \textit{response} of the expected answer.
    \item \textbf{Protocol Instruction}: We create a dataset to produce the protocol workflows defined in telecom standards following a human instruction. For example, an \textit{instruction} of ``Identify the different types of APN/DNN supportedm for UE implementation capabilities.", the expected \textit{response} created from standard is ``1. Verify the default APN/DNN setup; 2. Validate APN IDs for different purposes; 3. Align APN/DNN configurations; 4. Submit APN/DNN IDs". 
\end{itemize}

We further conduct model and human validation over the created telecom instruct following datasets. This includes rephrasing or removing ambiguous information referring context in the original pre-train data or not in the \textit{input} filed.

\begin{figure}
\begin{mdframed}[linecolor=black, linewidth=2pt, roundcorner=10pt]
\textbf{Prompt Template for Telecom Instruction Following}
\hrule height 0.5pt
\vspace{0.15in}
\textbf{Instruction:} You are specialized in Telecommunication domain. You are familiar with topics like 5G, RAN, wireless communication, etc., as well as technical Telecom standards, specifications from 3GPP. You are given a text in a 3GPP document. Your task is to transform it into an instruction as follows: \{``instruction": ``...", ``output": ``..."\}. Instruction types could be planning subtasks according to the 3GPP procedure to complete the main tasks or achieve a goal. The instruction/output must be clear. They must contain relevant context/passages in the target text needed to interpret them. Ensure that they do not mention implicit information, such as figures, tables, annexes, other sections in the document, etc.

\vspace{0.1in}
\textbf{Example}: \{``instruction": ``Initiate event based charging with decentralized and centralized unit determination, centralized rating", ``output": ``1. Request for resource usage. 2. Units Determination. 3. Charging Data Request. 4. Account, Rating Control. 5. Create CDR. 6. Charging Data Response. 7. Granted Units Supervision. 8. Content/Service Delivery."\}

\vspace{0.1in}
\textbf{TEXT}:
$\left \{ \text{text} \right\}$
\end{mdframed}
\caption{Prompt template for creating task completion or planning. Clear instructions are given to \acp{llm} to avoid references, implicit contents and guide \acp{llm} to generate sequential steps.}
\label{fig4:mcq_template}
\end{figure}

\section{Training Detail}
\label{sec:training}

Due to the constraint of available GPU resources, we limit the model size of base model to be less or equal than 8B. Our selected base model includes Llama2-7B, Mistral-7B and Llama3-8B and some corresponding instruct models.

\textbf{Domain-Specific Continual Pretraining}: We continue pretrain a general-purpose \acp{llm} with the OpenTelecom dataset on 8 AWS ml.p4d.24xlarge instances with a training time of approximately 6 hours. For Llama3-8B, we choose to not further pre-train it for two main reasons: i) the pre-trained dataset of Llama-3 series contains 15 TB tokens, our pre-training dataset is largely built on open documents on the web, it is uncertain to determine if our dataset is already used to train Llama3-8B; ii) our hardware limit us to efficiently continue pre-train in a reasonable batch size.

\textbf{Instruct Tuning}: We use QLoRA \cite{dettmers2024qlora} to instruct tune a pre-trained \ac{llm}. The reasons for selecting QLoRA are as follows: i) QLoRA is applied as a regularizer to avoid the catastrophic forgetting of fine-tuned \acp{llm}; ii) Due to the rapid advance of the telecom domain, our framework should keep flexible to fit the potential update of new standards and research achievements. The rank of the adapter is selected as $r = 512$ with $\alpha = 256$ and a dropout rate of $0.05$. For the training setting, we train the model with a constant learning rate $2.0\times 10^{-4}$ for $3$ epochs with AdamW optimizer \cite{loshchilov2017decoupled} without weight decay. Furthermore, we use FSDP \cite{Zhao2023PyTorchFSDP} to speed up the train with 8 GPUs with 32GB memory on one ml.p4d.24xlarge instance. The whole training took approximately $1.5$ hours.

\textbf{Alignment Tuning}: Similar to instruction tuning stage, we utilize QLoRA to regularize the model so that the aligned model is not deviated too far from the \ac{sft} model. We utilize an AdamW optimizer with a cosine learning rate scheduler with $5.0\times 10^{-6}$ as an initial learning rate. The rank of QLoRA adapter is $r=256$ with $\alpha=128$ with a dropout rate $0.1$. For DPO setting, $\beta=0.1$ is selected. We train the SFT model for one full epoch with FSDP as well.

In what follows, We will the models obtained after continual pretraining, instruct tuning and alignment tuning with the suffixes ``TP," ``TI," and ``TA", respectively. Finally, the best model obtained by combining different stages of our training pipeline, i.e., continual pre-training, instruct tuning and alignment tuning with DPO will be referred as TelecomGPT.  As our simulation will show, the best model obtained is actually a fine-tuned Llama3-8B with instruct tuning then followed by alignment tuning.

\section{Results}
\label{sec:results}
In this section, we demonstrate the performance of our telecom-specific \ac{llm} for multiple benchmarks. Qualitative examples of different benchmarks can be found in appendices.

\subsection{Traning Performance}

\begin{figure}[t!]
\centering
\includegraphics[width=1\linewidth]{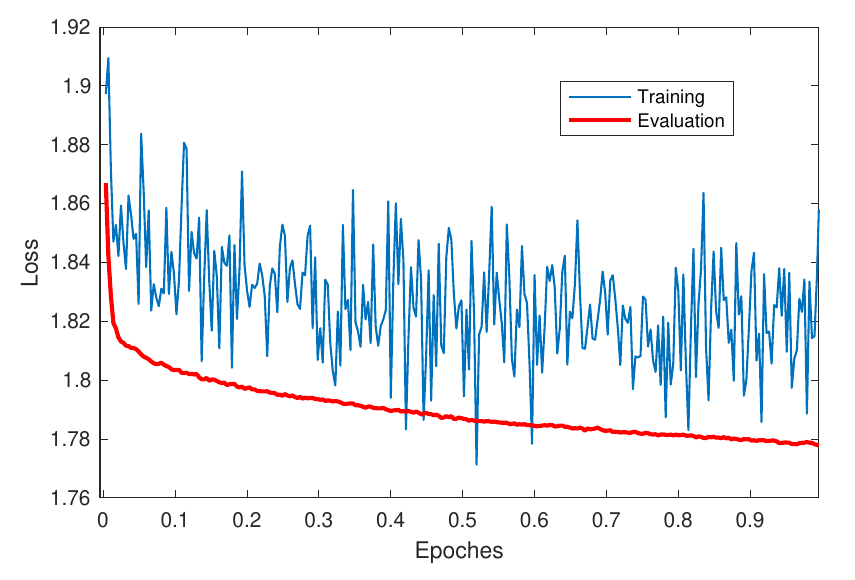}
\caption{Training and evaluation loss during continue pretraining (LlaMA2-7B-TP).}
\label{fig:pretraining}
\end{figure}

\textbf{Continual Pretraining}: Fig. \ref{fig:pretraining} shows the training and evaluation loss during continual pre-training on LlaMA2-7B. The evaluation loss first drop significantly then saturates gradually, suggesting a better domain adaptation in Telecom.

\begin{figure}[t!]
\centering
\includegraphics[width=1\linewidth]{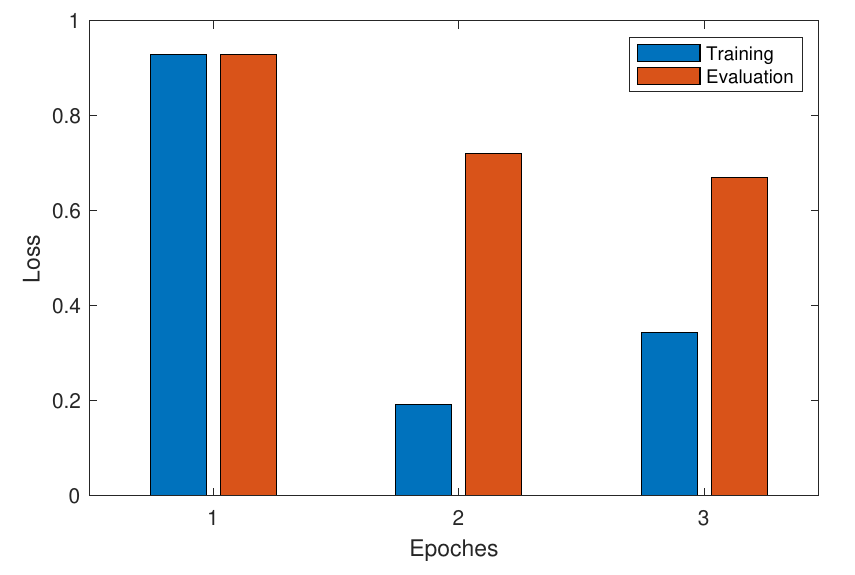}
\caption{Training and evaluation loss during instruct tuning (LlaMA3-8B-TI).}
\label{fig:instruct_tuning}
\end{figure}

\textbf{Instruct Tuning}: Fig. \ref{fig:instruct_tuning} shows the training and evaluation loss during instruct tuning on LlaMA3-8B. The notable evaluation loss shows that base model gradually learn all types of instructions in our instruction dataset. 

\begin{figure}[t!]
\centering
\includegraphics[width=1\linewidth]{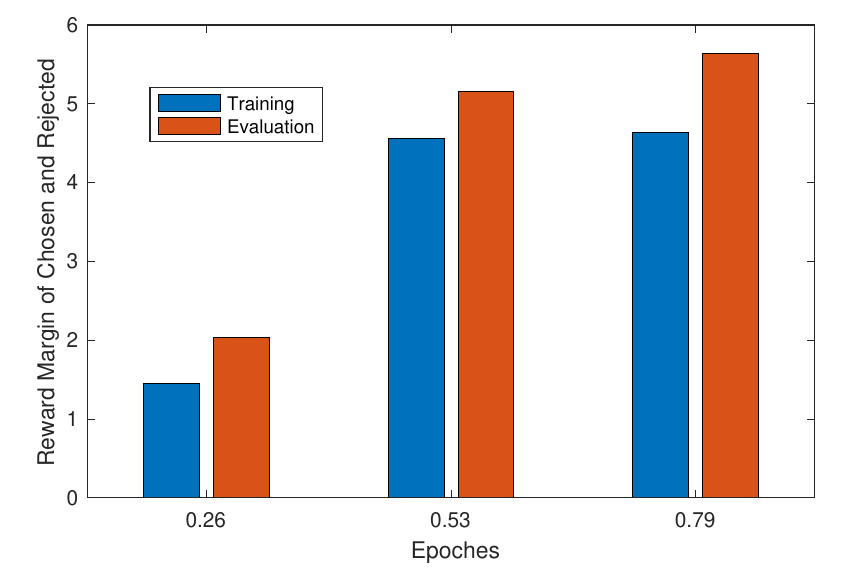}
\caption{Reward margin of chosen and rejected answer during alignment tuning (LlaMA3-8B-TI-TA).}
\label{fig:alignment_tuning}
\end{figure}

\textbf{Alignment Tuning}: Fig. \ref{fig:alignment_tuning} shows the reward margin (difference of reward) of chosen and rejected answers during alignment tuning on LlaMA3-8B, demonstrating clearly a tendency that responses similar to chosen samples are preferred than those similar to rejected samples.

\subsection{Evaluation Benchmark Performance}

\textbf{Telecom \ac{mcq}}:
In table \ref{tab:TeleQnA}, the accuracies of SOTA \acp{llm} and our \acp{llm} in different training stage are demonstrated in different categories of TeleQnA benchmark.  After instruct tuning of our \textbf{Telecom Instruct} dataset, there is an obvious increase of accuracy for all base models. Moreover, our telecom-specific models outpeforms the instruct tuned version of the base models and achieve a comparable performance compare to GPT-4 while the model size is notably smaller.  After the alignment tuning stage, there is a slight accuracy drop as well. This decrease might be caused by the selection strategy of alignment dataset (both correct and wrong option follow the same format of response). Moreover, for each category, there exist models with significant accuracy increase showing the impact of fine-tuning. Nevertheless, the average accuracy for standard specifications is still limited for all tested \acp{llm} since questions related to standard specifications contains techinical details such as the precise value of important qualities, the procedure of a defined scenario. For Llama2-7B, before instruct tuning, we add the one more continual pretraining stage. The effectiveness of continual pretraining is confirmed by an increase of accuracy by approximately $4\%$.
\begin{table*}[t!]
\centering
{%
\begin{tabular}{|>{\centering\arraybackslash}m{3cm}|c|c|c|c|c|c|}
\hline 
\acp{llm} & Lexicon & \makecell{Research \\ Overview} & \makecell{Research \\ Publications} & \makecell{Standards \\ Overview}  & \makecell{Standards \\ Specifications} & Overall \\
\hline \hline
GPT-4o & 92 & 81.73 & 79.54 & 83.87 & 62.77 & 78 \\ 
GPT-4 & 92 & 77 & 78 & 79 & 60 & 75 \\ 
GPT-3.5 & 96 & 66.35 & 66.98 & 64.52 & 56.38 & 66 \\ 
\hline
Llama3-8B & 72 & 51.92 & 65.11 & 56.45 & 36.17 & 56.20 \\ 
Llama3-8B-Instruct & 80 & 67.31 & 69.77 & 59.68 & 50 & 64.80 \\ 
\textbf{Llama3-8B-TI} & \textbf{96} & \textbf{69.23} & \textbf{74.88} & \textbf{74.19} & \textbf{56.38} &  \textbf{71.20} \\ 
{Llama3-8B-TI-TA} & 92 & 73.08 & 71.63 & 72.58 & 58.51 & 70.60 \\ 
\hline
Mistral-7B & 72 & 49.04 & 51.16 & 50 & 34.04 & 48.40 \\ 
Mistral-7B-Instruct & 84 & 64 & 65 & 56 & 51 & 62 \\ 
\textbf{Mistral-7B-TI} & \textbf{84} & \textbf{67.3} & \textbf{70.69} & \textbf{56.45} & \textbf{51.06} & \textbf{65.2} \\ 
{Mistral-7B-TI-TA} & {84}  & {70.19}  & {73.95} & {61.29}  & {48.94}  & {64} \\ 
\hline
LlaMA-2-7B & 62.5 & 52.24 & 49.18 & 48.28 & 40 & 48.94 \\ 
\textbf{LlaMA-2-7B-TI} & \textbf{84} & \textbf{57.69} & \textbf{63.26} & \textbf{56.45} & \textbf{50} & \textbf{59.80} \\ 
\textbf{LlaMA-2-7B-TP-TI} & \textbf{81.82} & \textbf{63.92} & \textbf{67} & \textbf{70} & \textbf{47.48} & \textbf{63.79} \\ 
\hline

\end{tabular}%
}
\caption{Performance Comparison of Telecom \ac{mcq} (TeleQnA) of mainstream \acp{llm}} and different telecom-specific \acp{llm} obtained during different stages of our training pipeline. \label{tab:TeleQnA}
\end{table*}

\begin{table}[t!]
\centering
\begin{tabular}{|>{\centering\arraybackslash}m{2cm}|c|c|c|c|c|c|c|}
\hline 
\acp{llm} &  Paper &  Book  & Patent & Wiki & Overall  \\
\hline \hline
GPT-4o & 98 & 94  & 84 & 98 & 93 \\ 
GPT-3.5 & 95 & 89  & 83 & 91 & 89 \\ 
\hline
LlaMA3-8B-Instruct & 97.92 & 89.81  & 80.65 & 87.21 & 88.49 \\ 
\textbf{LlaMA3-8B-TI} & \textbf{92.52} & \textbf{97.17} & \textbf{92.16}  & \textbf{94.12} & \textbf{94}  \\ 
\hline
{Mistral-7B-Instruct} & {78.5} & {76.42} & {65.69}  & {71.64} & {73.25} \\ 
\textbf{Mistral-7B-TI} & \textbf{80.37} & \textbf{82.08} & \textbf{82.35}  & \textbf{81.18} & \textbf{81.5} \\ 
\hline
\end{tabular}
\caption{Performance Comparison of Telecom \ac{mcq} on extended dataset} \label{tab:TeleQnAExt}
\end{table}

\textbf{Telecom Standard Classification}: Table  \ref{tab:3GPP_WG_Classification} summarizes the average accuracy of classifying 3GP Tdocs using \ac{sota} \acp{llm} and our telecom-specific \acp{llm} on a test dataset with 2000 texts of all 16 working groups. With only 1000 data samples per working group used during instruct tuning stage, most telecom-specific \acp{llm} achieve an average accuracy near 70$\%$ while the accuracy of GPT-4o and GPT-4 is less than 40$\%$. Moreover, the accuracy of classification for each TSG is slightly different (the accuracy of RAN is significantly higher than SA), demonstrating an imbalanced understanding of the 3GPP \acp{tdoc}.

\begin{table}[t!]
\centering
\begin{tabular}{|>{\centering\arraybackslash}m{2cm}|c|c|c|c|c|c|}
\hline 
\acp{llm} &  RAN &  SA  & CT & Overall  \\
\hline \hline
GPT-4o & 44.12 & 47.59  & 17.28 & 38.94 \\ 
GPT-3.5 & 42.93  & 48.59  & 16.26 & 38.54 \\ 
\hline
LlaMA3-8B-Instruct & 39.13 & 38.66 & 16.19 & 33.35 \\ 
\textbf{LlaMA3-8B-TI} & \textbf{82.76} & \textbf{68.8}  & \textbf{73.61}  & \textbf{75.30} \\ 
\hline
Mistral-7B-Instruct & 29.59 & 33.20 & 16.87  & 27.84 \\ 
\textbf{Mistral-7B-TI} & \textbf{76.62} & \textbf{76.90} & \textbf{49.89}  &  \textbf{70.83} \\ 
\hline
\end{tabular}
\caption{Performance Comparison of Telecom \ac{tdoc} Classification Benchmark of SOTA \acp{llm}} and our models \label{tab:3GPP_WG_Classification}
\end{table}

\begin{figure}[!ht]
\centering
\includegraphics[width=1\linewidth]{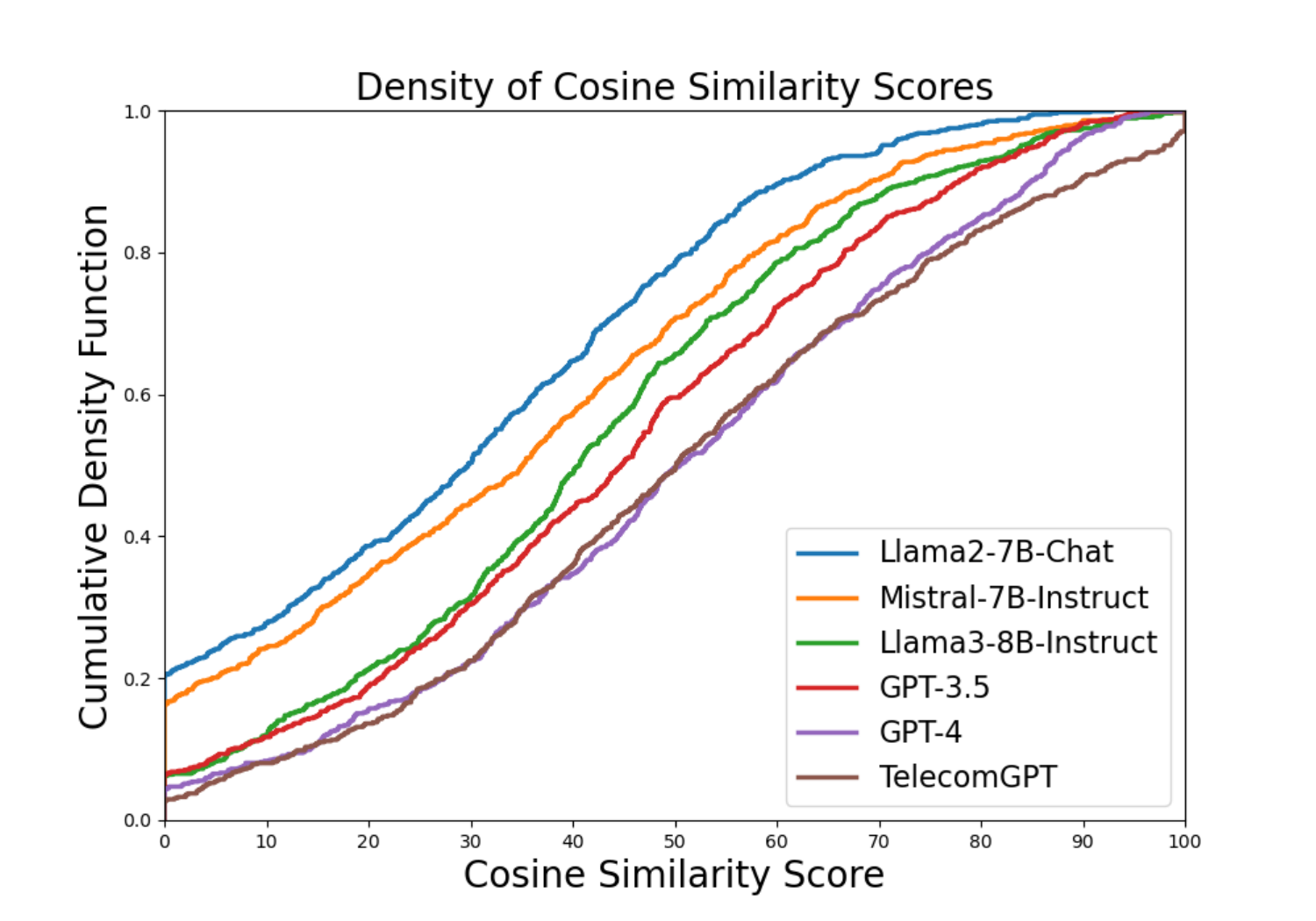}
\caption{Cumulative Density Function v.s. MathBERT score of Telecom Math Modeling Benchmark for SOTA \acp{llm} and our \acp{llm} obtained during different stages of training. Compared to GPT-4, our telecom-specific \ac{llm} achieve a better portion of high quality telecom-relevant equations.}
\label{fig:accuracy_vs_portion}
\end{figure}

\textbf{Telecom Math Modeling}: we compare the performance of mainstream \acp{llm} and our telecom-specific \acp{llm} obtained during different training stages here. In Table \ref{tab:TeleEq}, the average MathBERT score and the portion of predicted equations with MathBERT score greater than $90\%$ and $50\%$ are illustrated. Due to the difficulty of the task and the long context general required, we test the equation infilling capabilities on around 600 equations randomly extracted from 170 research papers which is never seen during training stage. One can observe that after SFT, the average MathBERT score outperform all mainstream \acp{llm} such as Llama3-8B-Instruct and Mistral-7B-Instruct with similar model size and GPT-3.5 and Mistral-8x7B with significant large model size. Besides, after SFT, our model achives a comparable portion of $50\%$ and a significant increase of portion of equations with $90\%$ representing that the predicted equation have almost the same operation tree as the ground truth and might be different in details such as operators, exponent, order of operations and etc.  The increase of average score after SFT shows that our telecom-specific instruction dataset. Moreover, after the alignment with human preference with the DPO algorithm, we observe a slight increase of score outperforming significantly SOTA \acp{llm} such as GPT-4 and GPT-4o with extremely larger model size and amount of pre-training data. Besides, the portion of predicted equations with score higher than $50\%$ is improved to be slightly better than SOTA \acp{llm}. The \acf{cdf} of the MathBERT score is illustrated in Fig. \ref{fig:accuracy_vs_portion}. The \ac{cdf} curve of TelecomGPT outperforms most \ac{sota} models and dominates the one of GPT-4 starting from a MathBERT score larger than $60\%$, showing the best math modeling capabilities when provided with clear context of the system model.  Finally, the average score for all \acp{llm} are close to $50\%$ entails the math modeling capabilities of current models are still relatively weak which demonstrates the hardness of proposed math modeling benchmark. 

\begin{table}[t!]
\centering
\begin{tabular}{|>{\centering\arraybackslash}m{2.5cm}|c|c|c|c|c|c|}
\hline 
\acp{llm} &  Average Score &   $\geq 90 \%$ &  $\geq 50 \%$   \\
\hline \hline
GPT-4 & 49.38 & 3.77  & 50.35 \\ 
GPT-3.5 & 43.53 & 1.81  & 40.44 \\ 
\hline
Llama3-8B-Instruct & 40.78 & 2.51 & 34.45 \\ 
Llama3-8B-TI & 46.16 &  {9.69}  & 46.80  \\ 
\textbf{Llama3-8B-TI-TA} &  \textbf{49.45} & \textbf{9.52} &  \textbf{50.73} \\ 
\hline
Mistral-7B-Instruct & 35.54 & 1.53 & 29.43  \\ 
Mistral-7B-TI & 47.66 & 8.04 & 48.77  \\ 
\textbf{Mistral-7B-TI-TA} & \textbf{48.11} & \textbf{7.22} & \textbf{49.26}  \\ 
\hline
\end{tabular}
\caption{Performance Comparison of Telecom Math Modeling Benchmark of SOTA \acp{llm}} \label{tab:TeleEq}
\end{table}

\textbf{Telecom Code Benchmark}: In Table \ref{tab:code}, we compare the performance of \acp{llm} for four code tasks proposed in our benchmark. Compared to general-purpose \acp{llm}, the \ac{sft} models and the aligned models both possess significantly higher Rouge scores which demonstrate an obvious improvement on the relevance of telecom contents. For code summary, analysis the Rouge scores of \ac{sft} models and aligned models are close. However, as our examples demonstrated in appendices, responses of the aligned models follow the concise style of the ground truth which is always selected as the ``chosen" sample of the \ac{dpo} algorithm. Besides, the responses provided by general-purpose \acp{llm} are detailed comprehensive but complicated (even irrelevant sometimes). Despite the fact that our telecom-specific models still hallucinates for in some requests, obtained telecom-specific \acp{llm} align with the preference we desire. 
\begin{table*}
\centering
{%
\begin{tabular}{ccccccccc}
\toprule
\multirow{2}{*}{\centering \acp{llm}} & \multicolumn{2}{c}{Code Summary} & \multicolumn{2}{c}{Code Analysis} & \multicolumn{2}{c}{Code Infilling} & \multicolumn{2}{c}{Code Generation} \\
\cmidrule(lr){2-3} \cmidrule(lr){4-5} \cmidrule(lr){6-7} \cmidrule(lr){8-9}
 & Rouge1 & RougeL & Rouge1 & RougeL & Rouge1 & RougeL & Rouge1 & RougeL \\
\midrule
Llama-3-8B-Instruct & 0.3202 & 0.2120 & 0.3336 & 0.2092 & 0.1125 & 0.0974 & 0.1914 & 0.1247 \\
{Llama3-8B-TI} & {0.5192} &  {0.3772} & {0.3533} & {0.2378} & {0.4169} & {0.3743} & {0.2255} & {0.1730} \\
\textbf{Llama3-8B-TI-TA} & \textbf{0.5167} & \textbf{0.3706} & \textbf{0.4048} & \textbf{0.2720} & \textbf{0.4336} & \textbf{0.3903} & \textbf{0.2712} & \textbf{0.2022} \\
Mistral-7B-Instruct & 0.3639 & 0.2342 & 0.3979 & 0.2411 & 0.1058 & 0.0882 & 0.1938 & 0.1264 \\
Mistral-7B-TI & 0.5701 & 0.3954 & 0.5459 & 0.3601 & 0.4347 & 0.3867 & 0.2513 & 0.1959 \\
\textbf{Mistral-7B-TI-TA} & \textbf{0.5751} & \textbf{0.3912} & \textbf{0.5473} & \textbf{0.3547} & \textbf{0.4394} & \textbf{0.3885} & \textbf{0.2602} & \textbf{0.2001} \\
\bottomrule
\end{tabular}}
\caption{Average F-measure of Rouge1 and RougeL scores for different code tasks in telecom domain.}\label{tab:code}
\end{table*}

\textbf{Telecom \ac{qa} and instruct following}: Table \ref{tab:task} presents the Rouge scores of different \acp{llm} with TelecomGPT in the tasks of Telecom domain open-ended \ac{qa}, general and protocol instruct following. The observation similar to code tasks still holds for these tasks. The instruct tuning significantly improves the Rouge scores of Llama-3 and Mistral in all tasks, and the alignment tuning improves open-ended \ac{qa} and general instruction tasks. Examples from Table \ref{tab:general_instruction_example_patent_1} to \ref{tab:general_instruction_example_3gpp_2} show that instruct tuning can make \ac{llm}'s responses more precise and concise compared to the references in patents, research papers, and standards. Furthermore, the alignment tuning eliminates the confusion of \ac{llm} in responding to general and protocol instructions. Moreover, results from Table \ref{tab:teleqna_standard_overview_1} to \ref{tab:teleqna_research_publications_2} show the fine-tuning stages make \ac{llm}'s answer better aligned with preferred answers. 

\begin{table*}
\centering
{%
\begin{tabular}{ccccccc}
\toprule
\multirow{2}{*}{\centering \acp{llm}} & \multicolumn{2}{c}{Open-ended \ac{qa}} & \multicolumn{2}{c}{General Instruction} & \multicolumn{2}{c}{Protocol Instruction}  \\ 
\cmidrule(lr){2-3} \cmidrule(lr){4-5} \cmidrule(lr){6-7} 
 & Rouge1 & RougeL & Rouge1 & RougeL & Rouge1 & RougeL \\
\midrule
Llama-3-8B-Instruct  & 0.0552 & 0.0496 & 0.1881 & 0.1430 & 0.2039 & 0.1303 \\
Llama-3-8B-TI & 0.3339 & 0.3233 & 0.3819 & 0.2908 & 0.3505 & 0.2449  \\
\textbf{Llama-3-8B-TI-TA}  & \textbf{0.4168} &  \textbf{0.4054} & \textbf{0.4142} & \textbf{0.3142}  & \textbf{0.3530} & \textbf{0.2438} \\
Mistral-7B-Instruct & 0.0595 & 0.0509 & 0.2055 & 0.1504 & 0.2188 & 0.1342  \\
Mistral-7B-TI & 0.2287 & 0.2190 & 0.4189 & 0.3219 & 0.3529 & 0.2483  \\
\textbf{Mistral-7B-TI-TA} & \textbf{0.2826} & \textbf{0.2695} & \textbf{0.4661} & \textbf{0.3651} & \textbf{0.3629} & \textbf{0.2375}  \\
\bottomrule
\end{tabular}}
\caption{Average F-measure of Rouge1 and RougeL scores of open \ac{qa} and instruct following.}\label{tab:task}
\end{table*}

\section{Conclusion}
\label{sec:conclusion}
In this paper, we propose a full pipeline to build telecom-specific \acp{llm}. Our data processing pipeline can effectively select telecom-relevant contents from general pretraining dataset. Then we can build instruction dataset and preference dataset to further instruct tune \acp{llm} to perform various downstream tasks and align with human preference. Besides, effective benchmarks to evaluate \acp{llm}'s capabilities in telecom context were proposed and verified with experimental results. Models obtained by telecom-specific fine-tuning pipeline outperform the corresponding base models and the start of the art models with significantly larger model size. Due to the resource limits, our experiments remain in small scale.  Pre-training from scratch with even larger dataset looks promising to build 
better telecom-specific \ac{llm}.  Moreover, our models can only treat textual data which limit the potential of current framework. One promising future direction is to integrate other data modalities, especially radio signal.
\bibliographystyle{IEEEtran}
\bibliography{IEEEabrv,Refs}



\begin{table*}[t!]
    \centering
    \begin{tabular}{lp{13cm}}
        \toprule
        \textbf{Prompt} & Describe the process of providing authorization to a mobile device for a pending transaction request.  \\ 
        \midrule
        \textbf{Category} & Patent \\
        \midrule
        \textbf{Ground Truth} & To provide authorization to a mobile device for a pending transaction request, the system searches the database for the mobile device's phone number, generates and sends an authorization request message to the mobile device, and awaits a reply message indicating approval or denial of the transaction from the user.  \\
        \midrule
        \textbf{Llama3-8B-Instruct} & When a mobile device receives a pending transaction request, the process of providing authorization involves several steps to ensure the security and integrity of the transaction. Here's an overview of the process: 1. **Transaction Request**: The merchant's server sends a transaction request to the mobile device, which includes the transaction details such as the amount, merchant name, and transaction type (e.g., purchase, payment, etc.). 2. **Authentication**: The mobile device checks if the user is authenticated and logged in to the device. If not, the user is prompted to log in or authenticate using biometric authentication (e.g., fingerprint, facial recognition), password, or PIN. 3. **Transaction Verification**: The mobile device verifies the transaction request by checking the transaction details, such as the amount, merchant name, and transaction type, to ensure it's a legitimate request. 4. **Authorization Request**: The mobile device sends an authorization request to the user, which includes the transaction details and a prompt to authorize or decline the transaction. 5. **User Response**: The user reviews the transaction details and responds to the authorization request by selecting "Authorize" or "Decline". 6. **Authorization Response**: The mobile device sends the user's response (authorize or decline) back to the merchant's server. 7. **Transaction Completion**: If the user authorizes the transaction, the merchant's server completes the transaction and sends a confirmation to the mobile device. If the user declines the transaction, the transaction is cancelled, and the merchant's server is notified. 8. **Transaction Logging**: The mobile device logs the transaction, including the transaction details, user response, and outcome (authorized or declined), for auditing and security purposes. Throughout this process, the mobile device ensures that the transaction is secure by using encryption, digital signatures, and other security measures to protect sensitive information. Additionally, the mobile device may also perform additional checks, such as verifying the user's account balance or available funds, before authorizing the transaction. \\
        \midrule
        \textbf{Llama3-8B-TI} & 1. The user of the mobile device receives a notification about a pending transaction request. 
        
        2. The user provides a PIN or biometric to the mobile device to authorize the transaction. 
        
        3. The mobile device communicates the authorization to the payment network, which then communicates the authorization to the merchant. \\
        \midrule
        \textbf{Llama3-8B-TI-TA} & To provide authorization for a pending transaction request on a mobile device, the user must input an authorization code received on the mobile device or at a separate device associated with the mobile device within a specified time period, such as three minutes. If the authorization code is not input within the time period, the pending transaction request will be automatically declined. \\
        \bottomrule
    \end{tabular}
    \vspace{4mm}
    \caption{Comparison of generated response for instruction task of describing telecom specific process in the patent. After instruct and alignment tuning with telecom-specific data, our models provide concise and high-quality description of a telecom specific process compared to the general purpose \ac{llm}.} 
    \label{tab:general_instruction_example_patent_1}
\end{table*}

\begin{table*}[t!]
    \centering
    \begin{tabular}{lp{13cm}}
        \toprule
        \textbf{Prompt} & Explain the difference between Shared VLAN Learning (SVL) database and Independent VLAN Learning (IVL) database designs in terms of distinguishing multiple VLANs for routing purposes.  \\ 
        \midrule
        \textbf{Category} & Patent \\
        \midrule
        \textbf{Ground Truth} & The Shared VLAN Learning (SVL) database design does not distinguish among multiple VLANs in a set of VLANs for routing purposes, whereas the Independent VLAN Learning (IVL) database design distinguishes among VLANs for routing purposes.  \\
        \midrule
        \textbf{Llama3-8B-Instruct} & In a network, VLANs (Virtual Local Area Networks) are used to segment devices into separate broadcast domains. When it comes to routing, there are two common database designs used to manage VLANs: Shared VLAN Learning (SVL) and Independent VLAN Learning (IVL). 
        
        **Shared VLAN Learning (SVL) Database: ** In an SVL database, all VLANs share the same MAC-VLAN table. This means that when a device learns a MAC address, it is stored in the table along with the corresponding VLAN ID. When a packet is received, the switch checks the MAC-VLAN table to determine the VLAN ID and then forwards the packet accordingly. Pros: * Simplifies configuration and management * Reduces memory usage Cons: * Limited scalability, as the MAC-VLAN table can become large and unwieldy * Can lead to VLAN ID conflicts if multiple VLANs share the same MAC address **Independent VLAN Learning (IVL) Database: ** In an IVL database, each VLAN has its own separate MAC-VLAN table. This means that each VLAN has its own set of learned MAC addresses, and the tables are not shared between VLANs. Pros: * Scalability, as each VLAN has its own table and can handle a large number of MAC addresses * Reduced risk of VLAN ID conflicts. Cons: * More complex configuration and management * Requires more memory to store multiple tables In terms of distinguishing multiple VLANs for routing purposes, IVL databases are generally more effective. Since each VLAN has its own table, the switch can accurately determine the VLAN ID and forward packets accordingly. In contrast, SVL databases may require additional processing and lookup tables to determine the correct VLAN ID. In summary, SVL databases are simpler and more memory-efficient, but may struggle with scalability and VLAN ID conflicts. IVL databases are more scalable and reduce the risk of VLAN ID conflicts, but require more complex configuration and management. The choice between SVL and IVL ultimately depends on the specific needs and requirements of the network. \\
        \midrule
        \textbf{Llama3-8B-TI} & 1. Shared VLAN Learning (SVL) database design allows a switch to distinguish multiple VLANs for routing purposes by associating a single MAC address with multiple CVIDs in the database.
        
        2. Independent VLAN Learning (IVL) database design, on the other hand, allows a switch to associate a separate MAC address for each CVID in the database, providing a unique MAC address for each VLAN. \\
        \midrule
        \textbf{Llama3-8B-TI-TA} & In terms of distinguishing multiple VLANs for routing purposes, the Shared VLAN Learning (SVL) database design allows a single database to hold information about all ports participating in VLANs, while the Independent VLAN Learning (IVL) database design requires each port to maintain its own database for VLAN-to-MAC address mapping. \\
        \bottomrule
    \end{tabular}
    \vspace{4mm}
    \caption{Comparison of generated response for instruction task in telecom patent. After instruct and alignment tuning with telecom-specific data, our model provides concise and high-quality explanation of a telecom specific concept compared to the general purpose \ac{llm}.} 
    \label{tab:general_instruction_example_patent_2}
\end{table*}

\begin{table*}[t!]
    \centering
    \begin{tabular}{lp{13cm}}
        \toprule
        \textbf{Prompt} & Summarize the challenges and solutions in the field of CSI feedback in massive MIMO systems utilizing DL-based methods.  \\ 
        \midrule
        \textbf{Category} & Research Paper \\
        \midrule
        \textbf{Ground Truth} & The challenges in CSI feedback in massive MIMO systems include the vast CSI matrix, compressed sensing inefficiency, and lack of standard CSI dataset collection. DL-based solutions involve using end-to-end Auto-Encoder architecture for compression and reconstruction, as well as considering factors like sampling range and network complexity in practical deployment.  \\
        \midrule
        \textbf{Llama3-8B-Instruct} & A very specific and technical question! CSI (Channel State Information) feedback is a crucial aspect in massive MIMO (Multiple-Input Multiple-Output) systems, as it enables the base station to adapt to the changing wireless channel conditions. However, in massive MIMO systems, the number of antennas and users increases, making CSI feedback a challenging task. Deep Learning (DL) based methods have been proposed to address these challenges. Here's a summary: **Challenges:**1. **Feedback overhead**: With a large number of antennas and users, the amount of CSI feedback data increases exponentially, leading to significant overhead and potential interference. 2. **Feedback delay**: The delay in receiving CSI feedback can result in outdated channel information, affecting the performance of the system. 3. **Channel estimation errors**: The accuracy of channel estimation is crucial, but errors can occur due to limited feedback resources and noisy measurements. **Solutions using DL-based methods:**1. **Compressive sensing**: DL-based compressive sensing methods can reduce the feedback overhead by compressing the CSI data while preserving its essential information. 2. **Feedback reduction**: DL-based methods can learn to reduce the feedback data by identifying the most important CSI information and discarding the rest. 3. **Feedback prediction**: DL-based feedback prediction methods can predict the CSI feedback based on historical data, reducing the feedback delay and overhead. 4. **Channel estimation**: DL-based channel estimation methods can improve the accuracy of channel estimation by learning from the available CSI feedback data and other system information. 5. **Hybrid beamforming**: DL-based hybrid beamforming methods can optimize the beamforming weights by learning from the CSI feedback data and other system information, improving the system performance.
        
        Some popular DL-based methods for CSI feedback in massive MIMO systems include: 
        
        1. Convolutional Neural Networks (CNNs) 2. Recurrent Neural Networks (RNNs) 3. Long Short-Term Memory (LSTM) networks 4. Autoencoders 5. Generative Adversarial Networks (GANs)
        
        These DL-based methods have shown promising results in reducing the feedback overhead, delay, and errors, while improving the system performance in massive MIMO systems. \\
        \midrule
        \textbf{Llama3-8B-TI} & 1. Challenges in CSI feedback for massive MIMO systems include the need for accurate channel estimation, efficient compression methods, and the ability to handle distributed data. 
        
        2. Solutions involve DL-based methods for channel estimation, such as CNN and DNN approaches, as well as DL-based feedback strategies for efficient compression and handling of distributed data. \\
        \midrule
        \textbf{Llama3-8B-TI-TA} & Challenges in CSI feedback for massive MIMO systems using DL-based methods include the need for efficient compression strategies and robustness against channel variations. Solutions involve developing lightweight DL architectures for channel estimation and feedback, as well as exploring quantization techniques to reduce the overhead of transmitting CSI to the baseband unit. \\
        \bottomrule
    \end{tabular}
    \vspace{4mm}
    \caption{Comparison of generated response for instruction task in telecom research paper. After instruct and alignment tuning with telecom-specific data, our model provides concise and high-quality summaries of telecom techniques compared to the general purpose \ac{llm}.} 
    \label{tab:general_instruction_example_arxiv_2}
\end{table*}

\begin{table*}[t!]
    \centering
    \begin{tabular}{lp{13cm}}
        \toprule
        \textbf{Prompt} & Explain the purpose of Test Tolerances in the context of 3GPP radiated tests defined for FR2, and how they affect the minimum requirements for testing.  \\ 
        \midrule
        \textbf{Category} & 3GPP \\
        \midrule
        \textbf{Ground Truth} & Test Tolerances in the context of 3GPP radiated tests defined for FR2 are used to relax the minimum requirements to create test requirements when a test requirement differs from the corresponding minimum requirement. The Test Tolerance for a test is non-zero, and it explains how the minimum requirement has been relaxed for that specific test by the Test Tolerance.  \\
        \midrule
        \textbf{Llama3-8B-Instruct} & In the context of 3GPP radiated tests defined for FR2 (Frequency Range 2), test tolerances play a crucial role in ensuring the reliability and consistency of test results. Test tolerances are the allowed variations in the measurement results, which account for the inherent uncertainties and limitations of the testing equipment, measurement procedures, and environmental conditions. The purpose of test tolerances is to: 1. Define the acceptable range of measurement uncertainty: By specifying the test tolerances, the 3GPP standards ensure that the measurement results are within a certain range of the actual values, taking into account the limitations of the testing equipment and procedures. 2. Ensure consistency and comparability: Test tolerances enable the comparison of measurement results from different testing laboratories, equipment, and procedures, by providing a common framework for evaluating the performance of devices under test (DUTs). 3. Allow for reasonable variations in testing conditions: Test tolerances account for variations in environmental conditions, such as temperature, humidity, and noise, which can affect the measurement results. In the context of FR2 radiated tests, test tolerances are particularly important due to the following reasons: 1. Increased complexity: FR2 tests involve more complex measurement procedures and equipment, which can introduce additional sources of uncertainty. 2. Higher frequency: FR2 operates at higher frequencies (24.25 GHz to 52.6 GHz), which can be more challenging to measure accurately due to the increased noise and interference. 3. Increased sensitivity: FR2 devices are designed to operate at higher frequencies, which requires more precise measurements to ensure compliance with the standards. The test tolerances defined in the 3GPP standards for FR2 radiated tests typically include: 1. Frequency tolerance: The allowed deviation from the nominal frequency. 2. Power tolerance: The allowed deviation from the nominal power level. 3. Phase tolerance: The allowed deviation from the nominal phase angle. 4. Amplitude tolerance: The allowed deviation from the nominal amplitude. By considering these test tolerances, the minimum requirements for testing are affected in the following ways: 1. Increased measurement uncertainty: The test tolerances account for the increased uncertainty in measurement results, which can lead to more conservative testing requirements. 2. More stringent testing conditions: To ensure compliance with the standards, testing conditions may need to be more stringent, such as using more precise equipment or more controlled environmental conditions. 3. Increased testing time: The test tolerances may require more testing time to ensure that the measurement results are within the allowed tolerances. \\
        \midrule
        \textbf{Llama3-8B-TI} & 3GPP radiated tests defined for FR2 include Test Tolerances that allow a test requirement to be relaxed by a specific value, resulting in a Test Requirement after Tolerance (TRaT). The minimum requirements for testing are derived by applying Test Tolerances to the nominal requirements, ensuring that the TRaT is achievable by a wide range of UE types. \\
        \midrule
        \textbf{Llama3-8B-TI-TA} & 3GPP radiated tests defined for FR2 specify Test Tolerances that allow relaxing the minimum requirements when testing. These tolerances are based on the measurement uncertainty of the test method, ensuring that the minimum requirements can be tested successfully within the specified limits. \\
        \bottomrule
    \end{tabular}
    \vspace{4mm}
    \caption{Comparison of generated response for instruction task in telecom standard. After instruct and alignment tuning with telecom-specific data, our model provides precise concise and high-quality explanation of telecom standard features compared to the general purpose \ac{llm}.} 
    \label{tab:general_instruction_example_3gpp_1}
\end{table*}

\begin{table*}[t!]
    \centering
    \begin{tabular}{lp{13cm}}
        \toprule
        \textbf{Prompt} & Identify the relationship between the channel bandwidth specified for NR V2X licensed band in Rel-16 and the potential introduction of new channel bandwidths for NR V2X licensed bands in Rel-17.  \\ 
        \midrule
        \textbf{Category} & 3GPP \\
        \midrule
        \textbf{Ground Truth} & The channel bandwidths defined for NR V2X licensed band in Rel-16 should be a subset of UE channel bandwidths for the same licensed band. If operators request new channel bandwidths for NR V2X licensed bands, it can be considered in the enhanced sidelink WI for defining additional RF requirements in Rel-17.  \\
        \midrule
        \textbf{Llama3-8B-Instruct} & A technical question! In Rel-16, the channel bandwidth for NR V2X (Vehicle-to-Everything) licensed band is specified as 10 MHz, 20 MHz, and 40 MHz. This is defined in 3GPP TS 38.104. For Rel-17, there is a potential introduction of new channel bandwidths for NR V2X licensed bands. Specifically, 3GPP is considering the addition of 80 MHz and 160 MHz channel bandwidths for NR V2X in the licensed bands. The relationship between the channel bandwidth specified in Rel-16 and the potential introduction of new channel bandwidths in Rel-17 is that the new bandwidths will provide improved spectral efficiency and capacity for NR V2X applications. The increased bandwidths will enable more devices to be supported, and higher data rates can be achieved, which is beneficial for applications such as vehicle-to-vehicle (V2V) and vehicle-to-infrastructure (V2I) communication. In summary, the introduction of new channel bandwidths in Rel-17 builds upon the foundation established in Rel-16, providing a more robust and efficient NR V2X system that can support a wider range of use cases and applications. \\
        \midrule
        \textbf{Llama3-8B-TI} & 1. The channel bandwidth specified for NR V2X licensed band in Rel-16 is 100 MHz. 2. The potential introduction of new channel bandwidths for NR V2X licensed bands in Rel-17 is 20 MHz and 40 MHz, in addition to the existing 100 MHz. \\
        \midrule
        \textbf{Llama3-8B-TI-TA} & The channel bandwidth specified for NR V2X licensed band in Rel-16 is aligned with existing NR FR1 bands defined in Rel-15. The potential introduction of new channel bandwidths for NR V2X licensed bands in Rel-17 is being considered to support additional use cases and requirements.\\
        \bottomrule
    \end{tabular}
    \vspace{2mm}
    \caption{Comparison of generated response for instruction task in telecom standard. After instruct and alignment tuning with telecom-specific data, our model provides precise concise and high-quality explanation of telecom standard features compared to the general purpose \ac{llm}.} 
    \label{tab:general_instruction_example_3gpp_2}
\end{table*}


\begin{table*}[t!]
    \centering
    \begin{tabular}{lp{13cm}}
        \toprule
        \textbf{Prompt} & Which version of SNMP provides authentication and privacy? [TCP/IP]  \\ 
        \midrule
        \textbf{Category} & Standards overview [TeleQnA] \\
        \midrule
        \textbf{Ground Truth} & SNMPv3  \\
        \midrule
        \textbf{Llama3-8B-Instruct} & SNMPv3 (Version 3) provides authentication and privacy. It was introduced in 2002 and is the latest version of the SNMP protocol. SNMPv3 includes features such as: 
        * Authentication: Ensures that the SNMP messages are coming from a trusted source and have not been tampered with.
        
        * Encryption: Encrypts the SNMP messages to prevent eavesdropping and unauthorized access. 
        
        * Message integrity: Verifies that the SNMP messages have not been altered during transmission. SNMPv3 is considered a more secure version of SNMP compared to earlier versions, which did not have built-in authentication and encryption mechanisms. \\
        \midrule
        \textbf{Llama3-8B-TI} & 2  \\
        \midrule
        \textbf{Llama3-8B-TI-TA} & SNMPv3\\
        \bottomrule
    \end{tabular}
    \vspace{2mm}
    \caption{Comparison of generated answers for questions in telecom domain. 
    After DPO, the inaccurate answer of instruct tuned model is corrected.}
    \label{tab:teleqna_standard_overview_1}
\end{table*}

\vspace{6mm}

\begin{table*}[t!]
    \centering
    \begin{tabular}{lp{13cm}}
        \toprule
        \textbf{Prompt} & What is the definition of Active communication?  \\ 
        \midrule
        \textbf{Category} & Lexicon [TeleQnA] \\
        \midrule
        \textbf{Ground Truth} & A UE is in active communication when it has a CS connection established.  \\
        \midrule
        \textbf{Llama3-8B-Instruct} & In telecommunications, active communication refers to a type of communication where the sender and receiver are actively involved in the communication process. This means that both parties are actively transmitting and receiving information, and are able to respond to each other in real-time. In other words, active communication is a two-way communication process where both parties are actively participating, sending and receiving information, and are able to respond to each other's messages. This is in contrast to passive communication, where one party is simply receiving information without actively participating in the communication process. Examples of active communication include:
        
        * Real-time voice or video calls
        
        * Instant messaging or chat sessions
        
        * Online conferencing or webinars
        
        * Interactive online forums or discussion boards
        
        Active communication is often used in situations where immediate feedback or response is required, such as in business meetings, customer service, or emergency situations. It allows for a more dynamic and interactive exchange of information, and can help to build stronger relationships and improve communication outcomes. \\
        \midrule
        \textbf{Llama3-8B-TI} & 1. A PLMN where the UE has found a cell that satisfies certain conditions for active communication.
        
        2. A PLMN where the UE has found a cell that satisfies certain conditions for camping on a cell. \\
        \midrule
        \textbf{Llama3-8B-TI-TA} & A UE is in active communication when it has a CS connection established or when it has one or more Activated PDP contexts. \\
        \bottomrule
    \end{tabular}
    \vspace{4mm}
    \caption{Comparison of generated answers for questions in telecom domain. After instruct tuning with telecom-specific tasks, the answer generated by our model provides precise answer compared to the general purpose \ac{llm}. After DPO, TelecomGPT is capable of providing concise and relatively accurate answer.}
    \label{tab:teleqna_research_overview_1}
\end{table*}

\begin{table*}[t!]
    \centering
    \begin{tabular}{lp{13cm}}
        \toprule
        \textbf{Prompt} & What is a broadcast channel?  \\ 
        \midrule
        \textbf{Category} & Research publications [TeleQnA] \\
        \midrule
        \textbf{Ground Truth} & A channel where multiple receivers communicate with a common sender  \\
        \midrule
        \textbf{Llama3-8B-Instruct} & In telecommunications, a broadcast channel is a communication channel that allows a single sender to transmit information to multiple receivers simultaneously. In other words, it's a one-to-many communication channel. In a broadcast channel, the sender (also known as the broadcaster) transmits a signal or data to a large number of receivers, such as a television station broadcasting a program to a wide audience, a radio station broadcasting music to listeners, or a mobile network broadcasting a message to multiple mobile devices. The key characteristics of a broadcast channel are: 
        
        1. One-to-many communication: The sender transmits information to multiple receivers. 
        
        2. Unidirectional transmission: The sender transmits information in one direction, from the sender to the receivers. 
        
        3. No feedback: The receivers do not send information back to the sender. 
        
        4. Shared medium: The broadcast channel uses a shared medium, such as the airwaves, to transmit the information. 
        
        Examples of broadcast channels include: 
        
        * Television broadcasting: A TV station broadcasts a program to a wide audience. 
        
        * Radio broadcasting: A radio station broadcasts music or talk shows to listeners. 
        
        * Mobile broadcasting: A mobile network broadcasts a message or data to multiple mobile devices. 
        
        * Satellite broadcasting: A satellite broadcasts a signal to a wide audience, such as a satellite TV channel.
        
        In contrast, a point-to-point channel is a communication channel that allows two devices to communicate with each other directly, whereas a broadcast channel allows a single sender to transmit information to multiple receivers. \\
        \midrule
        \textbf{Llama3-8B-TI} & A channel where one transmitter communicates with multiple receivers  \\
        \midrule
        \textbf{Llama3-8B-TI-TA} & A channel where one transmitter communicates with multiple receivers\\
        \bottomrule
    \end{tabular}
    \vspace{4mm}
    \caption{Comparison of generated answers for questions in telecom domain. 
    The answer provided by the base model is wordy while both our instruct tuned model and alignment tuned model provide the same answer which is accurate and concise.
    }
    \label{tab:teleqna_research_publications_1}
\end{table*}

\begin{table*}[t!]
    \centering
    \begin{tabular}{lp{13cm}}
        \toprule
        \textbf{Prompt} & What is a network slice according to 3GPP?  \\ 
        \midrule
        \textbf{Category} & Research publications [TeleQnA] \\
        \midrule
        \textbf{Ground Truth} & A logical network with specific characteristics and capabilities  \\
        \midrule
        \textbf{Llama3-8B-Instruct} & According to 3GPP (3rd Generation Partnership Project), a network slice is a logical partitioning of a network infrastructure to provide a specific set of network functions and resources to support a particular service or application. It is a key concept in 5G networks, enabling multiple independent networks to coexist on the same physical infrastructure. A network slice is a customized and isolated environment that provides a specific set of network functions, such as radio access, core network, and edge computing, tailored to meet the requirements of a particular service or application. This allows multiple services or applications to share the same physical infrastructure, while maintaining isolation and ensuring that each service or application receives the necessary resources and network functions to operate effectively. Network slices can be created and managed using software-defined networking (SDN) and network function virtualization (NFV) technologies, which enable the creation of virtual networks and network functions that can be dynamically provisioned and managed. In 3GPP, network slices are defined as a set of network functions and resources that are provisioned and managed as a single entity, and are designed to support specific use cases, such as: 
        
        * IoT (Internet of Things) 
        
        * Mission-critical communications 
        
        * Ultra-reliable low-latency communications (URLLC) 
        
        * Massive machine-type communications (mMTC) 
        
        * Enhanced mobile broadband (eMBB) 
        
        By providing a flexible and customizable network infrastructure, network slices enable service providers to offer a wide range of services and applications on the same physical infrastructure, while also reducing costs and improving network efficiency. \\
        \midrule
        \textbf{Llama3-8B-TI} & A set of virtual network functions (VNFs) that run on cloud infrastructure. \\
        \midrule
        \textbf{Llama3-8B-TI-TA} & A logical network built on virtualization technology to provide specific services\\
        \bottomrule
    \end{tabular}
    \vspace{4mm}
    \caption{Comparison of generated answers for questions in telecom domain. 
    There is a repeated definition on network slices in the response of the base model while the telecom-specific models provide two equivalent concise answers.
    }
    \label{tab:teleqna_research_publications_2}
\end{table*}



\begin{table*}[t!]
\centering
\begin{tabular}{lp{11.5cm}}
\toprule
\textbf{Prompt} & Please fill in the missing mathematical equation represented by $<\text{MASK}>$ placeholder in the provided sections of a telecom-relevant research paper that discuss the system model or problem formulation.  
\vspace{0.2in}

Network model: We consider two-dimensional wireless networks with the following communication nodes: A network of $L_T$ cooperating transmitters, a single receiver, 
and a network of $L_E$ passive eavesdroppers.
The passive eavesdroppers do not transmit any signal, and try to intercept the information that is transmitted between the pairs of legitimate
nodes, hence reducing the secrecy capability of the network.
Their locations are unknown to the transmitters.
Each transmitter is equipped with only a single omni-directional antenna.

In the sequel, we use the following notation:
\begin{description}
\item [$L(A)$] The area of a region $A\in R^2$;
\item [$L_T$] A random variable which denotes the number of transmitters in a region $A$;
\item [$L_E$] A random variable which denotes the number of eavesdroppers in a region $A$;
\item [$b||c$] A concatenation of two data blocks $b$ and $c$;
\item[$V, V_e$] The additive noise at receiver and eavesdropper, which are independent zero mean Gaussian random variables with variance $\sigma^2$;
\item[$C_{t,r}$] Capacity of the communication channel between transmitter $t$ and receiver $r$;
\item[$C_{s:t,r}$] Secrecy capacity between transmitter $t$ and receiver $r$;
\item[$C_s$] Secrecy capacity between a set of cooperating transmitters and a receiver;
\item[$d_{j,i}$] The distance between nodes $i$ and $j$.
\end{description}

We use the additive white Gaussian noise model.
Then, the received signal at the receiver $r$ from the transmitter $t$ is
\begin{eqnarray*} Y = d_{t,r}^{-\beta/2}X+V.\end{eqnarray*}

where $X$ is the transmitted signal from the transmitter $t$, and $\beta$ is the path-loss coefficient [Rappaport:book].
The received signal at the eavesdropper $e$ from the transmitter $t$ equals
\[ <\text{MASK}> \]

The point to point capacities between transmitter $t$ and receiver $r$, and between transmitter $t$ and eavesdropper $e$ are given by [Hellman:1978]. \\
\midrule
\textbf{Ground Truth} &  
\begin{eqnarray*}
Z_e = d_{t,e}^{-\beta/2}X+V_e.
\end{eqnarray*} 
\vspace{-0.1in}\\
\midrule
\textbf{Llama3-8B-Instruct} & \begin{eqnarray*} C_e = d_{t,e}^{-\beta/2}X + V_e \quad \text{[93.23\%]} \end{eqnarray*} 
\vspace{-0.15in}\\
\midrule
\textbf{Llama3-8B-TI} & \begin{eqnarray*} Y_e = d_{t,e}^{-\beta/2}X+V_e. \quad \text{[95.66\%]} \end{eqnarray*}   
\vspace{-0.15in} \\
\midrule
\textbf{Llama3-8B-TI-TA} &  \begin{eqnarray*} Z = d_{t,e}^{-\beta/2}X+V_e.   \quad \text{[93.59\%]} \end{eqnarray*}  
\vspace{-0.15in} \\
\bottomrule
\end{tabular}
\vspace{2mm}
\caption{Comparison of math modeling for system model extracted from \cite{Stojanovski2013SecureWC}. 
Our telecom-specfic \acp{llm} provide an accurate math modeling of the received signal (notice that no notation is predefined.) while general purpose \acp{llm} fail to follow the notation introduced earlier.}
\label{tab:telecom_equation_example_1}
\end{table*}

\begin{table*}[t!]
\centering
\begin{tabular}{lp{11.5cm}} %
\toprule
\textbf{Prompt} & Please fill in the missing mathematical equation represented by $<\text{MASK}>$ placeholder in the provided sections of a telecom-relevant research paper that discuss the system model or problem formulation.  
\vspace{0.1in}

\textbf{System Model}: We consider a discrete-time zero-mean jointly proper Gaussian flat-fading channel with the following input-output relation
\vspace{-0.05in}
\[ <\text{MASK}> \]
\vspace{-0.05in}
with the diagonal matrices $\mathbf{H}=\textrm{diag}(\mathbf{h})$ and $\mathbf{X}=\textrm{diag}(\mathbf{x})$. Here the $\textrm{diag}(\cdot)$ operator generates a diagonal matrix whose diagonal elements are given by the argument vector. The vector $\mathbf{y}=\left[y_{1},\hdots,y_{N}\right]^{T}$ contains the channel output symbols in temporal order. Analogous, $\mathbf{x}=\left[x_{1},\hdots,x_{N}\right]^{T}$, $\mathbf{n}=\left[n_{1},\hdots,n_{N}\right]^{T}$, and $\mathbf{h}=\left[h_{1},\hdots,h_{N}\right]^{T}$ contain the channel input symbols, the additive noise samples and the channel fading weights. All vectors are of length $N$. The samples of the additive noise process are assumed to be i.i.d.\ zero-mean jointly proper Gaussian with variance $\sigma_{n}^{2}$ and, thus, $\mathbf{R}_{n}=\mathrm{E}\left[\mathbf{n}\mathbf{n}^{H}\right]=\sigma_{n}^{2}\mathbf{I}_{N}$, with $\mathbf{I}_{N}$ being the identity matrix of size $N\times N$.
The channel fading process is zero-mean jointly proper Gaussian with the temporal correlation characterized by
\[ <\text{MASK}> \]
Its variance is given by $r_{h}(0)=\sigma_{h}^{2}$. For mathematical reasons we assume that the autocorrelation function $r_{h}(l)$ is absolutely summable, i.e., 
\[ <\text{MASK}>\]
\vspace{-0.2in}
\\
\midrule
\vspace{-0.1in}
\textbf{Ground Truth} &  
\begin{equation*}
\mathbf{y} =\mathbf{H}\mathbf{x}+\mathbf{n}=\mathbf{X}\mathbf{h}+\mathbf{n}   
\end{equation*}
\begin{equation*}
r_{h}(l) =\mathrm{E}[h_{k+l}\cdot h_{k}^{*}].
\end{equation*}
\begin{equation*}
\sum_{l=-\infty}^{\infty}|r_{h}(l)|<\infty. 
\end{equation*}

\\
\midrule
\textbf{Llama3-8B-Instruct} & \begin{equation*}
\mathbf{y}[n] = \mathbf{H}\mathbf{x}[n] \quad \text{[70.72\%]} 
\end{equation*}
\begin{equation*}
\mathbf{R}_h = \sigma_h^2 \mathbf{I}_N + (\mathbf{R}_h \otimes \mathbf{R}_h)^{1/2} \quad \text{[48.33\%]} 
\end{equation*}
\begin{equation*}
\sum_{l=-\infty}^{\infty}|r_h(l)|<\infty
\quad \text{[97.65\%]} 
\end{equation*} 
\vspace{-0.1in}  \\
\midrule
\textbf{Llama3-8B-TI} & 
\begin{equation*}
\mathbf{y} = \sqrt{\frac{P}{n_t}}\mathbf{H}\mathbf{x} + \mathbf{w} \quad \text{[54.48\%]} 
\end{equation*}
\begin{equation*}
\mathrm{E}\left[h_{i}h_{j}\right]=r_{h}(i-j)\:,\:\text{for}\:i\geq j\:. \quad \text{[54.94\%]} 
\end{equation*}
\begin{equation*}
\sum_{l=-\infty}^{\infty}r_{h}(l) < \infty.
\quad \text{[96.20\%]} 
\end{equation*} 
\vspace{-0.1in} \\
\midrule
\textbf{Llama3-8B-TI-TA} &  \begin{equation*}
\mathbf{y}=\sqrt{\gamma}\mathbf{H}\mathbf{x}+\mathbf{n}
\quad \text{[74.45\%]} 
\end{equation*}
\begin{equation*}
\mathrm{E}\left[h_{i}h_{j}^{*}\right]=r_{h}\left(\left|i-j\right|\right),\quad i\geq j.
\quad \text{[56.51\%]} 
\end{equation*}
\begin{equation*}
\sum_{l=-\infty}^{\infty}|r_{h}(l)| < \infty.
\quad \text{[99.32\%]} 
\end{equation*} 
\vspace{-0.1in} \\
\bottomrule
\end{tabular}
\vspace{2mm}
\caption{Comparison of math modeling for problem formulation extracted from \cite{Drpinghaus2011OnTG}. For better readability, we display three generated equations together while the equations are generated sequentially. Our telecom-specfic \acp{llm} provide a better math modeling of the channel while the general-purose \ac{llm} misses the channel noise and introduces a time-varying modeling which is not mentioned in the context.}
\label{tab:telecom_equation_example_2}
\end{table*}


\begin{table*}[t!]
\centering
\begin{tabular}{lp{12.5cm}}
\toprule
\textbf{Prompt} & Identify and infill the missing lines indicated by a placeholder $<\text{FILL}>$ required to complete this function's operation within its telecommunications context. \footnotesize{
\begin{verbatim}
Get MTU value in host OS.
    Args:
        iface(str):  Interface for getting MTU in host OS
        <fill>
        ifr = struct.pack('16sH', iface.encode("utf-8"), 0)
        mtu = struct.unpack('16sH', ioctl(soc, self.SIOCGIFMTU, ifr))[1]
    except Exception as err:
        raise PypackerException("ERROR: Getting MTU failed; {}".format(err))
    return mtu
\end{verbatim}
} \\
\midrule
\textbf{Ground Truth} &  \footnotesize{\begin{verbatim}
        Returns:
    int: Original MTU value
Examples::
    env.tg[1].get_os_mtu(iface=ports[('tg1', 'sw1')][1])
"""
try:
    soc = socket.socket(socket.AF_PACKET, socket.SOCK_RAW)
\end{verbatim} } \\
\midrule
\textbf{Llama3-8B-Instruct} & \footnotesize{\begin{verbatim}
import struct
import ioctl
soc = socket.socket(socket.AF_INET, socket.SOCK_DGRAM)
\end{verbatim} } \\
\midrule
\textbf{Llama3-8B-TI} &  \footnotesize{\begin{verbatim}
     Returns:
    int:  MTU value in host OS
Examples::
    env.tg[1].get_os_mtu(iface=ports[('tg1','sw1')][1])
"""
try:
    soc = socket.socket(socket.AF_PACKET, socket.SOCK_RAW)
\end{verbatim}} \\
\midrule
\textbf{Llama3-8B-TI-TA} &  \footnotesize{\begin{verbatim}
     Returns:
    int: OS MTU value
Examples::
    env.tg[1].get_os_mtu(iface=ports[('tg1','sw1')][1])
"""
soc = self._get_socket()
\end{verbatim}} \\
\bottomrule
\end{tabular}
\vspace{4mm}
\caption{Comparison of code infilling task of different \acp{llm}. The missing code lines predicted by our instruct tuned model is very close to the ground truth compared to the base model. Nonetheless, the response of the aligned model is slightly different from the instruct tuned model.}
\label{tab:code_infilling_examples}
\end{table*}

\begin{table*}[t!]
\centering
\begin{tabular}{lp{13.5cm}}
\toprule
\textbf{Prompt} & Analyze this script's contribution to telecom workflows, specifying its primary functionalities.  \footnotesize{
\begin{verbatim}
def bsc(input_bits, p_t):
    """
    Binary Symmetric Channel.

    Parameters
    ----------
    input_bits : 1D ndarray containing {0, 1}
        Input array of bits to the channel.

    p_t : float in [0, 1]
        Transition/Error probability of the channel.

    Returns
    -------
    output_bits : 1D ndarray containing {0, 1}
        Output bits from the channel.
    """
    output_bits = input_bits.copy()
    flip_locs = (random(len(output_bits)) <= p_t)
    output_bits[flip_locs] = 1 ^ output_bits[flip_locs]
    return output_bits
\end{verbatim}
} \\
\midrule
\textbf{Ground Truth} &  The script defines a function \texttt{bsc} for Binary Symmetric Channel within the telecom domain. It takes \texttt{input\_bits} array of bits and \texttt{p\_t} transition/error probability as parameters and returns \texttt{output\_bits} after applying the channel operations. The function simulates a binary symmetric channel by flipping bits based on the given probability, which is a crucial concept in telecom communication systems for modeling channel errors and noise. \\
\midrule
\textbf{Mistral-7B-Instruct} & The provided script is a Python function named \texttt{bsc} that simulates a Binary Symmetric Channel (BSC), a fundamental model in information theory used to study the effects of noise on digital communication channels.

The primary functionalities of this script are:

\begin{enumerate}
    \item It takes two parameters: \texttt{input\_bits}, an array of binary digits (0 or 1), and \texttt{p\_t}, the transition or error probability of the channel.

    \item It creates a copy of the input bits array, \texttt{output\_bits}.

    \item It generates a boolean array \texttt{flip\_locs} containing \texttt{True} or \texttt{False} values, where \texttt{True} indicates the location of the bits that will be flipped due to noise. The locations are determined by comparing each element of the range [0, length of \texttt{output\_bits}) with \texttt{p\_t}. If the comparison result is \texttt{True}, the corresponding index is included in \texttt{flip\_locs}.

    \item It flips the bits at the locations specified in \texttt{flip\_locs} by performing the XOR operation with 1. This effectively changes 0 to 1 and 1 to 0 at those locations, simulating the effect of noise.

    \item Finally, it returns the modified \texttt{output\_bits} array, which represents the output of the BSC.
\end{enumerate}

In the context of telecom workflows, this script can be used to analyze and optimize error correction codes, modulation schemes, and other aspects of digital communication systems by simulating the effects of noise on transmitted data. This can help in designing more reliable and efficient communication systems. \\
\midrule
\textbf{Mistral-7B-TI} & The script defines a function named bsc that simulates a binary symmetric channel. It takes input bits and the transition probability as parameters, and the function flips bits based on the transition probability. This functionality is relevant in the telecom domain for modeling and analyzing error correction mechanisms in digital communication systems. \\
\midrule
\textbf{Mistral-7B-TI-TA} & The script defines a function named bsc that simulates a binary symmetric channel. It takes input bits and an error probability parameter, and outputs the flipped bits based on the specified transition probability. This functionality is relevant in the telecom domain for modeling and analyzing the performance of binary transmission systems under error-inducing channels. \\
\bottomrule
\end{tabular}
\vspace{4mm}
\caption{Comparison of code summary task of different \acp{llm}. While the summary generated by the base model is indeed detailed, it does not consistently prioritize telecomm-related functionalities. In contrast, summaries generated by both the instruct-tuned and aligned model exhibit conciseness and maintain a consistent focus on telecom aspects.}
\label{tab:code_summary_examples}
\end{table*}

\begin{table*}[t!]
\centering
\begin{tabular}{lp{15.5cm}}
\toprule
\textbf{Prompt} & Develop a C function that retrieves and returns the transmit (TX) power value for a wireless device. The function should update a data structure representing wireless request (wrqu) with the TX power information including the power value, the fixed state, flags, and disabled status based on the device's status. This function is intended to be used in the context of configuring and managing wireless interfaces within a telecommunications network.   \\
\midrule
\textbf{Response} &  \footnotesize{
\begin{verbatim}
union iwreq_data *wrqu, char *extra)
{
    struct ipw_priv *priv = ieee80211_priv(dev);
    mutex_lock(&priv->mutex);
    wrqu->power.value = priv->tx_power;
    wrqu->power.fixed = 1;
    wrqu->power.flags = IW_TXPOW_DBM;
    wrqu->power.disabled = (priv->status & STATUS_RF_KILL_MASK) ? 1 : 0;
    mutex_unlock(&priv->mutex);
    IPW_DEBUG_WX("GET TX Power -> %s %d \n",
                 wrqu->power.disabled ? "OFF" : "ON", wrqu->power.value);
    return 0;
}
\end{verbatim}
}
\\
\midrule
\textbf{Mistral-7B-Instruct} & \footnotesize{
\begin{verbatim}
struct wrqu get_tx_power(int device_id) {
    struct wrqu wrqu;

    // Get the device status for the given device ID
    struct device_status device_status = get_device_status(device_id);

    // Assuming that the TX power and other details are stored in the device_status structure
    wrqu.power = device_status.tx_power;
    wrqu.fixed_state = device_status.fixed_state;
    wrqu.flags = device_status.flags;
    wrqu.disabled_status = device_status.disabled_status;

    return wrqu;
}
\end{verbatim}
} \\




\midrule
\textbf{Mistral-7B-TI-TA} & \footnotesize{
\begin{verbatim}
static int iwl_get_tx_power(struct iwl_priv *priv, struct iwl_wrqu *wrqu)
{
    struct iwl_tx_power_data tx_power;
    int err;
    err = iwl_dvm_get_tx_power(priv, &tx_power);
    if (err)
        return err;
    wrqu->data.length = sizeof(struct iwl_tx_power_data);
    wrqu->data.flags |= IWL_FLAG_FIXED;
    wrqu->data.fixed_state = tx_power.fixed;
    wrqu->data.flags |= tx_power.flags;
    wrqu->data.disabled = tx_power.disabled;
    wrqu->data.value = tx_power.value;

    return 0;
}
\end{verbatim}

}
\\
\bottomrule
\end{tabular}
\vspace{1mm}
\caption{Comparison of code generation of different \acp{llm}. The advantage of script generated by the base model is well commented while the telecom-specific models just generate pure code lines. This might is caused by the fact that commented functions are few in our datasets. Furthermore, the discrepancies between the ground-truth code and the response of telecom-specific \acp{llm} arise due to the fact the former is extracted from a complete code file with well-established context. Nonetheless, our telecom-specific \acp{llm} successfully encapsulate the vast majority of the intended functionalities, demonstrating their efficacy in telecom domain.}
\label{tab:code_generation_examples}
\end{table*}

\end{document}